\documentclass{ar-1col}
\usepackage[margin=2.5cm]{geometry}

\jname{Xxxx. Xxx. Xxx. Xxx.}
\jvol{AA}
\jyear{YYYY}
\doi{10.1146/((please add article doi))}

\usepackage{natbib}

\begin{document}

\markboth{Stanimirovi\'{c} \& Zweibel}{Atomic and Ionized Microstructures in the Diffuse ISM}

\def\hi{{\mbox{\sc H i}}}
\def\ion#1#2{\hbox{#1\,{\sc i}}} 
\def\arcdeg{\hbox{$^\circ$}}
\def\slantfrac#1#2{\hbox{$\,^#1\!/_#2$}}
\def\kms{km~s$^{-1}$}
\def\apss{Ap\&SS}
\def\degree{$^{\circ}$}
\def\nl{\cr}
\def\phm#1{\phantom{#1}}
\def\AAS{{\it A{\rm \&\/}AS\/}}
\def\ga{\mathrel{\hbox{\rlap{\hbox{\lower4pt\hbox{$\sim$}}}\hbox{$>$}}}}
\def\la{\mathrel{\hbox{\rlap{\hbox{\lower4pt\hbox{$\sim$}}}\hbox{$<$}}}}

\def\aap{\it Astron.~Astrophys.}
\def\aapr{\it Astron.~Astrophys.~Rev.}
\def\aaps{\it Astron.~Astrophys.~Suppl.}
\def\aas{\it Astron.~Astrophys.~Suppl.}
\def\actaa{\it Acta~Astron.}
\def\aip{\it AIP~Conf.~Ser.}
\def\aj{\it Astron.~J.}
\def\al{\it Astron.~Lett.}
\def\ao{\it Appl.~Opt.}
\def\apj{\it Ap.~J.}
\def\apjl{\it Ap.~J.~Lett.}
\def\apjs{\it Ap.~J.~Suppl.}
\def\apjsupp{\it Ap.~J.~Suppl.}
\def\aplett{\it Ap.~Lett.}
\def\applopt~{\it Appl.~Opt.}
\def\apspr{\it Ap.~Space~Phys.~Res.}
\def\apss{\it Ap.~Space~Sci.}
\def\araa{\it Annu.~Rev.~Astron.~Astrophys.}
\def\aspcs{\it ASP~Conf.~Ser.}
\def\ass{\it Ap.~Space~Sci.}
\def\astap{\it Astron.~Astrophys.}
\def\azh{\it Astron.~Zh.}
\def\baas{\it Bull.~Am.~Astron.~Soc.}
\def\bac{\it Bull.~Astron.~Inst.~Czechosl.}
\def\bain{\it Bull.~Astron.~Inst.~Neth.}
\def\caa{\it Chin.~Astron.~Astrophys.}
\def\cjaa{\it Chin.~J.~Astron.~Astrophys.}
\def\fcp{\it Fundam.~Cosmic~Phys.}
\def\gca{\it Geochim.~Cosmochim.~Acta}
\def\grl{\it Geophys.~Res.~Lett.}
\def\iau{\it IAU~Circ.}
\def\iaucirc{\it IAU~Circ.}
\def\icarus{\it Icarus}
\def\jcap{\it J.~Cosmol.~Astropart.~Phys.}
\def\jcp{\it J.~Chem.~Phys.}
\def\jgr{\it J.~Geophys.~Res.}
\def\jqsrt{\it J.~Quant.~Spec.~Radiat.~Transf.}
\def\jrasc{\it J.~R.~Astron.~Soc.~Can.}
\def\memras{\it MmRAS}
\def\memsai{\it Mem.~Soc.~Astron.~Ital.}
\def\mn{\it MNRAS}
\def\mnras{\it MNRAS}
\def\nar{\it New~Astron.~Rev.}
\def\nat{\it Nature}
\def\npa{\it Nucl.~Phys.~A}
\def\nphysa{\it Nucl.~Phys.~A}
\def\pasa{\it Publ.~Astron.~Soc.~Aust.}
\def\pasj{\it Publ.~Astron.~Soc.~Jpn.}
\def\pasp{\it Publ.~Astron.~Soc.~Pac.}
\def\physrep{\it Phys.~Rep.}
\def\physscr{\it Phys.~Scr.}
\def\planss{\it Planet.~Space~Sci.}
\def\pnas{\it PNAS}
\def\pr{\it Phys.~Rep.}
\def\pra{\it Phys.~Rev.~A}
\def\prb{\it Phys.~Rev.~B}
\def\prc{\it Phys.~Rev.~C}
\def\prd{\it Phys.~Rev.~D}
\def\prd{\it Phys.~Rev.~D}
\def\pre{\it Phys.~Rev.~E}
\def\prl{\it Phys.~Rev.~Lett.}
\def\procspie{\it Proc.~SPIE}
\def\qjras{\it Q.~J.~R.~Astron.~Soc.}
\def\rmaa{\it Rev.~Mex.~Astron.~Astrofis.}
\def\rmp{\it Rev.~Mod.~Phys.}
\def\rmxaa{\it Rev.~Mex.~Astron.~Astrofis.}
\def\sci{\it Science}
\def\skytel{\it Sky~Telesc.}\
\def\solphys{\it Sol.~Phys.}
\def\sovast{\it Sov.~Astron.}
\def\sp{\it Proc.~SPIE}
\def\ssr{\it Space~Sci.~Rev.}
\def\zap{\it Z.~Astrophys.}

\newcommand{\kapGP}{\kappa_f}
\newcommand{\kapP}{\kappa_P}
\newcommand{\kGP} {k_f}
\newcommand{\vGP} {v_f}
\newcommand{\kP}  {k_P}
\newcommand{\lAD} {\lambda_{AD}}
\newcommand{\lO}  {\lambda_{\Omega}}
\newcommand{\led}  {\lambda_{e}}
\newcommand{\lnum}  {\lambda_{num}}
\newcommand{\tAD} {\tau_{AD}}
\newcommand{\tO}  {\tau_{\Omega}}
\newcommand{\tGP} {\tau_f}
\newcommand{\cAi} {c_{Ai}}
\newcommand{\cAn} {c_{An}}
\newcommand{\RM} {R_{M}}
\newcommand{\RAD} {R_{AD}}
\newcommand{\mbfB}{\mathbf{B}}
\newcommand{\mbfF}{\mathbf{F}}
\newcommand{\mbfI}{\mathbf{I}}
\newcommand{\mbfJ}{\mathbf{J}}
\newcommand{\mbfP}{\mathbf{P}}
\newcommand{\mbfQ}{\mathbf{Q}}
\newcommand{\mbfb}{\mathbf{b}}
\newcommand{\mbfe}{\mathbf{e}}
\newcommand{\mbfg}{\mathbf{g}}
\newcommand{\mbfu}{\mathbf{u}}
\newcommand{\mbfui}{\mathbf{u}_i}
\newcommand{\mbfun}{\mathbf{u}_n}
\newcommand{\mbfuD}{\mathbf{u}_D}
\newcommand{\mbfuGP}{\mathbf{u}_{GP}}
\newcommand{\mbfz}{\mathbf{z}}
\newcommand{\mbfn}{\mathbf{n}}
\newcommand{\mbfk}{\mathbf{k}}
\newcommand{\mbfx}{\mathbf{x}}
\newcommand{\mbfy}{\mathbf{y}}
\newcommand{\nuin}{\nu_{in}}
\newcommand{\mbfnabla}{\mathbf{\nabla}}

\newcommand{\f}   {\frac}
\newcommand{\ddx}{\frac{\partial}{\partial x}}
\newcommand{\ddy}{\frac{\partial}{\partial y}}
\newcommand{\ddt}{\frac{\partial}{\partial t}}
\newcommand{\mq}{\langle q\rangle}
\newcommand{\dq}{\delta q}
\newcommand{\ppe}{P_{\perp}}
\newcommand{\ppa}{P_{\parallel}}
\newcommand{\ape}{\alpha_{\perp}}
\newcommand{\bnabla}{{\mbox{\boldmath$\nabla$}}}
\newcommand{\bkappa}{{\mbox{\boldmath$\kappa$}}}
\newcommand{\bxi}{{\mbox{\boldmath$\xi$}}}
\newcommand{\batea}{{\mbox{\boldmath$\eta$}}}


\title{Atomic and Ionized Microstructures in the Diffuse Interstellar Medium}

\author{Sne\v{z}ana Stanimirovi\'{c},$^1$ Ellen G. Zweibel,$^{1,2}$ }
\affil{$^1$Department of Astronomy, University of Wisconsin, Madison, USA, 53706; email: sstanimi@email.edu}
{\affil{$^2$Department of Physics, University of Wisconsin, Madison, USA, 53706}

\begin{abstract}
It has been known for half a century that the interstellar medium (ISM) of our Galaxy is structured on scales as small as a few hundred km, 
more than 10 orders of magnitude smaller than typical ISM structures and energy input scales.  In this review we focus on 
neutral and ionized structures 
on spatial scales of a few to $\sim10^4$ Astronomical Units (AU)
which appear to be highly overpressured, as these have the most important role in the dynamics and energy balance of interstellar gas: the Tiny Scale Atomic Structure (TSAS) and Extreme Scattering Events (ESEs) as the most over-pressured example of the Tiny Scale Ionized Structures (TSIS).
We review observational results and highlight key physical processes at AU scales. 
We present evidence for and against microstructures as part of a universal turbulent cascade and as discrete structures, and review their association with supernova remnants, the Local Bubble, and bright stars. We suggest a number of observational and theoretical programs that could clarify the nature of AU structures.  
TSAS and TSIS probe spatial scales in the range of what is expected for turbulent dissipation scales, therefore are of key importance for constraining exotic and not-well understood physical processes which have implications for many areas of astrophysics. The emerging picture is one in which a magnetized, turbulent cascade, driven hard by a local energy source and acting jointly with phenomena such as thermal instability, is the source of these microstructures.

\end{abstract}

\maketitle

\tableofcontents

{\it ``If there is the slightest foundation for these remarks the zoology of Archipelagoes 
will be well worth examining; for such facts (would) undermine the stability of Species." Charles Darwin, 1836}

\section{INTRODUCTION}

The diffuse interstellar medium (ISM) in the Galaxy contains structure over a wide range of spatial scales. For the neutral medium, a diverse hierarchy of structures on spatial scales $>1$ pc has been observed for decades, while the ionized medium has been known
to exhibit structure on much smaller spatial scales reaching down to a hundred of kilometers \citep{Rickett77,Armstrong95,Haverkorn13}. 

The cornerstone of most ISM models is a rough pressure equilibrium between different thermal phases \citep{Ferriere98}. The coexistence of the multiple phases, however, requires a well-defined, narrow range of thermal pressures $P/k \sim 3000$ K cm$^{-3}$ (see Section~\ref{s:pressure})\footnote{We use Gaussian cgs units throughout this paper.}.
Since the late 70s, however, sporadic observations of small-scale ISM structures with highly unusual properties have started to emerge. The Tiny Scale Atomic Structure (TSAS term first introduced by Heiles 1997)
) has been observed on spatial scales of a few to 10$^4$ Astronomical Units (AUs) using several different observational techniques at radio and optical/ultraviolet wavelengths, with observationally inferred hydrogen density of $n_H \sim10^4$ cm$^{-3}$. At temperatures of
60-260 K, TSAS appears hugely over-pressured with $P/k \sim 10^6$ K cm$^{-3}$. In the mid 80s the Tiny Scale Ionized Structure (TSIS) was discovered in radio observations of an extreme scattering event (ESE). With even smaller spatial scales, an electron density $n_e>100$ cm$^{-3}$ at $T>1000$ K again implied similarly large thermal pressures. A schematic summary of basic properties of TSIS and TSAS is shown in Figure~\ref{f:size-pressure}. Their implied thermal pressure is on average at least 100 times higher than the theoretically expected range for the multiphase medium ($P_{min}$ to $P_{max}$).

These observational results were so mind-boggling that Lyman Spitzer wrote in a letter to Phil Diamond on January 15, 1997: ``In view of the revolutionary importance which such structures have in our understanding of the interstellar gas, I am trying to understand the relevant observations and how they are interpreted.'' While the abundance of such over-pressured structures is still not well constrained, their story may be somewhat reminiscent of Charles Darwin's discovery of slightly different forms of mockingbirds and tortoises on the different islands of the Galapagos archipelago, or, to be mundane, the stray thread that when tugged unravels an entire garment. 

From the beginning, the study of TSAS and TSIS has been haunted by three interrelated issues: energetics, geometry, and pervasiveness. Attempts to explain the electron density fluctuations which cause pulsar scintillation, the first form of small scale structure discovered \citep{Scheuer68}, as compressive plasma waves ran into the difficulty that the waves
are strongly damped and require a large power source \citep{Scheuer68}. Although it was not commented upon at the time, the heating resulting from so much 
dissipation would also be large (\S~\ref{s:refraction-basics}).

With the discovery of TSAS (Dieter et al. 1976) and extreme scattering events (Fiedler et al. 1987), both hinting at significant overpressures and short relaxation times, the power source problem was exacerbated. 
At the same time, it was realized that both
the energetics and pressure problems can be alleviated if the density structures are 
highly elongated and viewed end on \citep{Romani87,Heiles97}. 
Identifying the physical processes that create such structures
and quantifying their effects and observational signatures has been a challenging problem.
%
Progress toward solving these  problems  
depends on knowing more about the overall environment in which small scale structure exists. Is it in one of 
the pervasive phases of the ISM, such as cold neutral or warm ionized gas? Is it an interface 
phenomenon that occurs in supernova remnant shells, bubble walls, cloud edges, or shocks? 
Or is there a population of tiny, self gravitating clouds, entities unto themselves?
Over time, explanations have coalesced around two poles, one favoring special, discrete structures and the other based on statistical properties of interstellar turbulence.

Understanding of TSAS and TSIS is important as the existence and properties of these structures could be undermining our understanding of heating, cooling and dynamical processes in the ISM, and potentially could lead to the discovery of physical processes that we have yet to account for. In addition,  
as pointed out by \citet{Smith13}, the existence of small and highly dense regions in the ISM has important implications for interstellar chemistry. At high densities the interstellar radiation field is attenuated, resulting in an enhanced abundance of molecular species. And if, as we discuss later, TSAS and TSIS are associated with strong dissipation, the resulting free energy could power unusual chemistry.

Besides addressing important aspects of the mainstream ISM, AU-scale structure could represent an important bridge to planetary science, and several recent studies have started to explore this avenue. \citet{Ray17} suggested that  hydroxyl radical (OH), neutral hydrogen (HI), and other spectral lines offer an exciting way to probe structure on $\sim1000$ AU scales around pulsars and 
search for the planet-building material.
Boyajian's star, an extraordinary star found by Planet Hunters whose properties appear to require TSAS to explain its flux variability, demonstrates the importance of TSAS for future planet searches \citep{Wright16,Simon17}.

While we focus in this review on the atomic and ionized microstructure within the Milky Way, observational evidence is starting to emerge about the AU-scale structure in external galaxies.
For example, \citet{Hacker13} studied absorption lines in quasar spectra using the Sloan Digital Sky Survey and
found evidence for absorption line variability in the intervening systems on scales 10-100 AU, suggesting that TSAS  is likely to exist in host foreground galaxies. \citet{Boisse15} found structure on similar scales in Mg II and Fe II temporal variability using Very Large Telescope and Keck observations.
Finally, there are some indications that even damped Lyman $\alpha$ absorbing systems at high redshift may contain HI structure on scales of $\sim10-100$ AU \citep{Kanekar01}.

In this review, we focus on TSAS and TSIS (mainly traced by extreme scattering events) as these microstructures have a clear over-pressure problem. 
AU-scale structure has also been observed in the diffuse molecular medium. Considering the intrinsically large pressure of molecular clouds due to self-gravity, magnetic, and surface forces, the over-pressure problem becomes less pronounced for AU-scale molecular structure. 

Although literature up through 2017 is cited, the bibliography is not comprehensive, and due to space limitations we had to omit discussion of many important papers. The reader is referred to excellent articles in the ASP Conf. Series Vol. 365
\citet{Haverkorn07} for complementary details  on optical absorption line studies, or the Galactic 
tiny-scale structure in the  ionized and molecular gas,  to
preceding ARAA articles on TSIS \citep{Rickett90,Rickett77}, and to the recent review by \cite{Haverkorn13}. 

\begin{figure}
\includegraphics[scale=0.7]{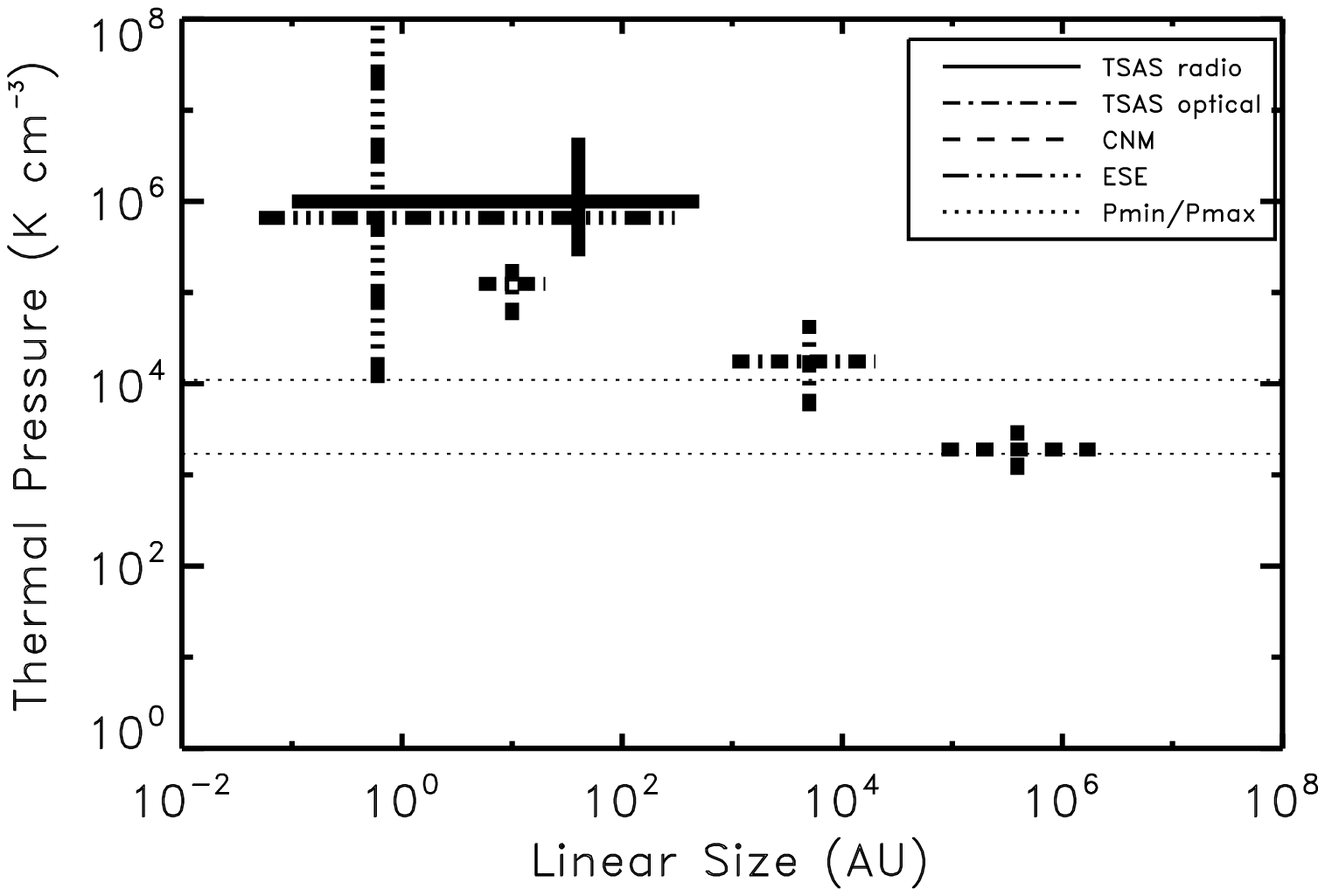}
\caption{A schematic summary of basic properties of TSAS and TSIS we consider in this review. The dotted lines show the rough range of allowed thermal pressures based on Wolfire et al. (2003) and Cox (2005). Properties of the Cold Neutral Medium (CNM) clouds are based on McKee \& Ostriker (1977) and \citet{Spitzer78}.  }
\label{f:size-pressure}
\end{figure}

\subsection{Fundamentals: The Phases of the Interstellar Medium}

The diffuse ISM does not uniformly populate
the density-temperature parameter space 
but
is known to exist in several aggregations or phases.
Two flavors of the neutral medium exist, corresponding to two phases:
the cold neutral medium (CNM) and the warm neutral
medium (WNM). Traditionally, the CNM and WNM are understood as being 
two thermal equilibrium states of the neutral medium 
\citep{Field69,McKee77a,Wolfire03}. 
The theoretically expected properties of the CNM and WNM, based on the heating and cooling balance, are: 
a kinetic temperature  $T_k\sim60-260$ K and a volume density 
of $n_H \sim7-70$ cm$^{-3}$ for the CNM\footnote{We note that 
CNM temperature can be even lower in the case when 
dust heating via photoelectric effect is not effective, \cite{Spitzer78}.
For example, if cosmic rays and X-rays are the only heating sources,
the CNM is expected to have temperature of $\sim20$ K (Mark Wolfire, private communication).},
and $T_k=5000-8300$ K and $n_H\sim0.2-0.9$ cm$^{-3}$ for the WNM 
(Wolfire et al. 2003, Table 3 for the Solar neighborhood). 
The diffuse warm ionized medium (WIM) has 
$n_H\sim 0.3$ cm$^{-3}$ and $T_k\sim8000$ K, while
the hot ionized medium (HIM) has 
$n_H\sim 3\times 10^{-3}$ cm$^{-3}$ and $T_k\sim 10^6$ K,
Draine (2011). The diffuse ISM phases have a roughly comparable thermal pressure \citep{Ferriere98}.
Molecular clouds and HII regions, on the other hand, are often not considered
as ISM phases due to their overpressure relative to the four phases.
While molecular clouds are gravitationally confined, HII regions are undergoing expansion.

While the ISM phases are traditionally understood as thermal equilibrium 
states, more recent ISM models and observations have emphasized 
the highly dynamic and turbulent character of the ISM and 
the consequences this has  on the fraction of gas outside of the equilibrium states.
For example, 
\citet{Audit05} showed that a collision of 
turbulent flows can initiate fast condensation of WNM into cold neutral clouds with 
the fraction of cold gas, as well as the fraction of thermally unstable gas,
being controlled by turbulence.
\citet{Koyama02} and \citet{MacLow05} explored the importance of 
shocks driven into warm, magnetized, and turbulent  gas
by supernova explosions. They found a continuum of gas temperatures, 
with a fraction of the thermally-unstable WNM being 
constrained by the star formation rate. The fractions of cold and thermally unstable
gas, however, vary widely with numerical models and can depend on input physics 
but also  numerical resolution \citep{Kim08}.
The question of whether the ISM phases are, on average, in thermal and dynamical equilibrium 
is still open from both theoretical and observational perspectives 
\citep[e.g.][]{Heiles03b,Begum10,Murray17} to cite just a few examples.

Two points should be borne in mind when applying the theory of thermal phases to small scale structure.  First, coexisting phases are separated by thin fronts in which thermal conduction is important, the temperature is unstable, and gas either evaporates or condenses, depending on pressure. We show in \S~\ref{s:neutral_theory} that TSAS is comparable in size to the thickness of these fronts. Second, although pressure fluctuations in a turbulent medium can mediate transitions between phases, the turnover time of a turbulent eddy decreases with
spatial scale such that there is a
minimum scale at which there is time for
heating and cooling processes to operate. Below this scale, the gas is approximately adiabatic. 

\subsection{Interstellar Thermal Pressure Fluctuations}
\label{s:pressure}

\begin{figure}
\includegraphics[scale=0.5]{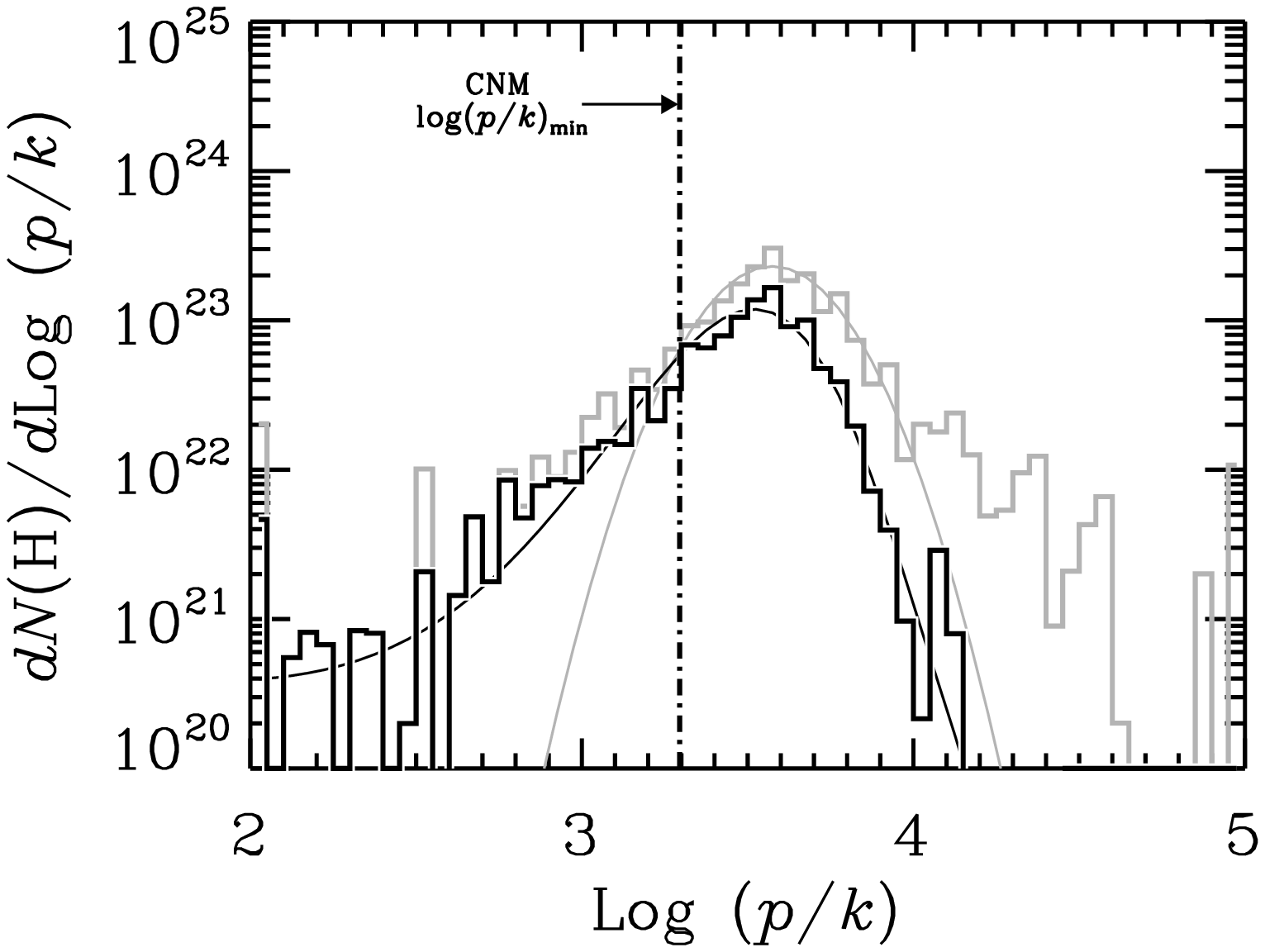}
\caption{The distribution of thermal pressure measurements from Jenkins \& Tripp (2011). Two distributions are shown: all measurements in gray, and a sub-sample with $I/I_0<10^{0.5}$ in black. The sub-sample is a better representation of the general ISM away from bright stars. Both distributions were weighted by the hydrogen column density. Lines show fitted expressions provided in Jenkins \& Tripp (2011).}
\label{f:pressure_observed}
\end{figure}

One of the main problems regarding TSAS is its (model-dependent) over-pressure relative to the
traditional thermal pressure of the ISM, which suggests that TSAS cannot persist nor be pervasive.  We summarize here briefly theoretical and observational pressure constraints.

The theoretical constraints regarding thermal pressure come from the observational finding
that the CNM and WNM coexist spatially. For this to happen, considering different heating and cooling processes,
requires a well-defined and narrow range of 
thermal pressures $P_{min}/k=2 \times 10^3$ to $P_{max}/k=5 \times 10^3$ K cm$^{-3}$ (Wolfire et al. 2003;
this range corresponds to the Solar neighborhood).
Dynamic and turbulent processes can increase the allowed pressure range, however the median 
thermal pressure of $P/k \sim3000$ cm$^{-3}$ is still a good representation \citep{Wolfire15}.

Several observational studies have measured thermal pressure.
\citet{Jenkins11} used ultraviolet spectra of 89 stars to identify three fine-structure lines of atomic carbon
C I. They found a median thermal pressure of $3800$ cm$^{-3}$, with the central portion of the pressure distribution being
fitted with a log-normal function. However, as shown in Figure~\ref{f:pressure_observed}, 
both low- (at $P/k <10^3$ K cm$^{-3}$) and high-end 
($P/k >10^{4}$ K cm$^{-3}$) portions of the  pressure distribution
showed significant populations. Although the formation
and maintenance of the high-pressure regions are open questions, Jenkins \& Tripp (2011) argued that such regions are found close to massive stars and are likely affected by dynamical processes like shocks and stellar winds. A small fraction ($\sim0.05$\% by mass) of their measurements have $P/k >10^{5.5}$ K cm$^{-3}$. 
Estimates of {\textit{total}} pressure, including magnetic fields, cosmic rays, and turbulence are of order $P/k=11,000 - 20,000$ K cm$^{-3}$  \citep{Cox05}, strongly suggesting that the high end of the Jenkins \& Tripp distribution cannot be in thermal equilibrium (especially in view of the lower thermal pressures calculated by Wolfire et al. 2003).
One possibility, suggested by several studies, is that TSAS traces the high-pressure regions revealed by C I. The under-pressure regions revealed by Jenkins \& Tripp  observations are also fascinating and potentially important for understanding the TSAS puzzle, we return to this point in \S~\ref{s:neutral_theory}. 

\citet{Goldsmith13} used ultraviolet CO lines toward nearby stars and found a slightly higher (average) thermal
pressure, $P/k=4600-6800$ K cm$^{-3}$, while \citet{Gerin15} used C II observations toward 13 sightlines in the
Galactic plane to measure $P/k=5900$ K cm$^{-3}$. Very recently, \citet{Herrera-Camus17} combined C II and HI
observations of nearby galaxies to estimate the C II cooling rate. Their results agree with Jenkins \& Tripp (2011), and also confirm the finding that thermal pressure increases with the interstellar radiation field and the star formation activity. 


As shown in the schematic figure (Figure 1),
according to most interpretations (with the exception of TSIS
that causes pulsar scintillation) the column densities and transverse dimensions of
both ESEs and TSAS imply very large internal pressure unless the line of
sight dimension is much longer than the transverse dimension. This requires either that
the structures be self confined or that they are transitory but replenished, in a statistical sense, by a powerful source.

The energy output by freely expanding small scale structures could be considerable. Suppose the structure consists of clouds that survive for time $\tau_s$, 
which for overpressured 
clouds of size $l$ and internal velocity dispersion $v$ would be the 
free expansion time $l/v$. If there are $n_c$ 
clouds per unit volume then the pathlength $\lambda$ that has to be observed for one detection 
is $\lambda\sim (\pi n_cl^2)^{-1}$. That is, if we observe $N_s$ sources at an average distance $D_s$ 
and detect $X$ small scale structure components, then $\lambda = N_sD_s/X$, and the volume filling factor $f\sim (4/3)l/\lambda$.

The expansion of a cloud of pressure $P_c$ injects energy at the rate $\dot E\sim 4\pi P_cl^2v\sim 4\pi P_cl^3/\tau_s$,
(assuming that there are no radiative losses). The energy 
input rate per volume is $n_c\dot E\sim 3P_cfv/l$. For an overpressure of 100, 
$P_c\sim 10^{-11}$ dyn cm$^{-2}$ (or $P_c/k \sim 7 \times 10^{4}$ K cm$^{-3}$). If $v\sim 1$ km s$^{-1}$ and $\lambda\sim 10$ AU, the energy input rate is 
about $2\times 10^{-20}f$ erg cm$^{-3}$ s$^{-1}$. In comparison, \cite{Wolfire03} fit the radiative cooling rate of the neutral ISM (their eqn. 3) by the formula $\Lambda(T)=1.4\times 10^{-27}T_2^{0.8}$ erg cm$^3$ s$^{-1}$, where $T_2\equiv T/10^2$K; this formula is accurate to 35\%  for $0.55 < T_2 < 85$ (and $n$ refers to the mean ambient density, not the TSAS density). The condition that heating by TSAS
expansion not exceed radiative cooling is then $f < 7\times 10^{-8}T_2^{0.8}n^2$. For
example, if $T_2 = 0.2-1$ and $n=100$ cm$^{-3}$, $f$ must be less than $(2-7) \times 10^{-4}$.
While 
these numbers are rough and not rigorous, it seems that on energetic grounds the ISM cannot tolerate too 
large a population of highly overpressured clouds.

\subsection{Turbulent Fluctuations in the Interstellar Medium}

Interstellar turbulence plays a very important role in the ISM, 
e.g. \citep{Miville-Deschenes10,Armstrong95}. Turbulence is far too broad a topic to be reviewed here; please see \cite{Elmegreen04,Scalo04,Lazarian2009} for fairly recent discussions of astrophysical
turbulence and a textbook such as \cite{Tennekes1972} for basic concepts. Here we only introduce
a few central ideas and terms that we will refer to later in the paper. Additional features of turbulence are discussed in \S 2.4.2.

A {\textit{turbulent 
cascade}} refers to motions over a range of spatial scales that are generated nonlinearly by energy input $\dot E$ at a particular scale (for example, a sound wave of wavenumber $k$ beats with
itself to create a sound wave of wavenumber $2k$). If the energy transfer between scales
conserves energy, that part of the cascade is called the {\textit{inertial range}}. 
Over the inertial
range, isotropic turbulence in incompressible gas follows the famous Kolmogorov scaling
law according to which $E_kdk$, the kinetic energy between wavenumber $k$ and $k+dk$, is
proportional to $k^{-5/3}$ (if we write the integral over wavenumber space in 3 dimensions
then $E_k=4\pi k^2P_k$, where the power spectrum $P_k\propto k^{-11/3}$). If a
strong magnetic field is present, motions parallel to it are suppressed relative to perpendicular motions but the perpendicular motions still follow Kolmogorov scaling. The contours of constant power in $\mbfk$ space are elongated, with $k_{\parallel}/k_{\perp}\sim (k_d/k_{\perp})^{1/3}$, where $k_d$ is the wavenumber at which the
turbulence is driven \citep{Goldreich95}. These properties can be derived by assuming energy
conservation together with the concept of ``critical balance", i.e. that the eddy turnover
rate $k_{\perp}v_{k,perp}$ is the same as the frequency of a parallel-propagating Alfven wave,
$k_{\parallel}v_A$. In the limit of extreme compressibility we have Burgers turbulence, which
is comprised entirely of shocks and has a $k^{-2}$ energy spectrum. When compressibility is accounted for, subsonic to transonic turbulence is accompanied by density fluctuations with the same power spectrum as the velocity fluctuations, but \cite{Kritsuk07} found a log normal density fluctuation spectrum in numerical simulations of isothermal hydrodynamic turbulence with Mach number 6.

If energy is
dissipated by a process such as viscosity which increases with decreasing scale, then the
inertial range terminates at the scale where the dissipation rate equals the energy transfer rate. For Kolmogorov turbulence this scale is $l_K\sim (\rho\nu^3/\dot E)^{1/4}$, where $\nu$
is the kinematic viscosity. The cascade at scales below the inertial range is called the
{\textit{dissipation range}}, and typically cuts off exponentially.

Many studies
have observed the turbulent spatial power spectrum down to spatial scales of $\la1$ pc using many different tracers \citep{Elmegreen04}. The slope of the 3D power spectrum depends on the interstellar tracer, optical depth, and velocity resolution.
For example, the electron density fluctuations follow a Kolmogorov spectrum over the range of scales from $\sim30$ pc to $\sim70$ km (Armstrong et al. 1995, see Section 3).
A 3D power spectrum slope of $-3.7$ was measured for the warm, optically-thin, Galactic HI seen in emission \citep{Dickey00}, or $\sim -3$ for HI in an optically-thick region close to the Galactic plane in agreement with theoretical expectations for the optically-thick medium \citep{Lazarian04}. Several carbon monoxide (CO) studies of even denser and optically-thick medium have found an even flatter slope of $-2.8$ \citep{Stutzki98}.
\citet{Miville-Deschenes10} derived a spatial power spectrum of dust column density fluctuations in the Polaris flare region over the range of scales 
from $\sim2000$ AU to $\sim20$ pc
and found a slope of $-2.7$.
It is expected that interstellar turbulence extends further down even to smaller spatial scales \citep{Hennebelle12}.
However, the nature of dissipation precesses, spatial scales on which they operate, and the local heating induced by turbulent dissipation, are complex questions that
still need to be constrained observationally. 

Several studies have hypothesized that TSAS could be related to the turbulent energy cascade on larger scales and possibly even trace the tail-end of the turbulent spectrum. For example,
\citet{Deshpande00a} suggested that the optical depth variations ascribed to TSAS are
primarily due to contributions from the large scale end of the interstellar turbulent cascade, thereby circumventing the overpressure problem.
We discuss this further in \S~\ref{s:neutral_theory}.

As mentioned above, turbulence in a strong magnetic field is anisotropic, with
elongated eddies. In estimating number densities from column density of TSAS and TSIS it is often assumed that the line of sight and transverse dimensions are similar. As pointed out by \cite{Heiles97}, the inferred densities and pressures would be lower if TSAS/TSIS were filaments viewed edge on. The connection to magnetized turbulence is not straightforward, however, because the cascade described above is incompressible.  The incompressible cascade does generate density fluctuations of an amplitude that scales with Mach number, but they are isotropic \citep{Cho03} and do not produce elongated structures. 
There is direct evidence for anisotropic TSIS structures \citep{Armstrong1981}. 
A recent study by \cite{Kalberla16}  found evidence for anisotropy in the HI spatial power spectrum for an intermediate latitude Galactic field. While the spectral index of the power spectrum remained close to Kolmogorov value, the spectral power changed with the position angle.


\section{Microstructures in the Neutral ISM
}
The Tiny-Scale Atomic Structure or TSAS  has been observed in the ISM for over four decades using
many different observational approaches. 
In the radio, spatial and temporal variability of HI absorption line profiles in the direction of background non-pulsar sources 
(summarized in Section~\ref{s:obsTSAS-VLBI}) have been used extensively.
Extragalactic compact and resolved, single or multiple, radio continuum
sources are commonly used as non-pulsar targets.
Several Galactic supernova remnants have been used as extended background sources to map out the distribution of the absorbing HI. One Galactic microquasar was also used as a target for temporal observations. 
Pulsars have several unique characteristics that make them especially exciting as targets for the temporal variability of absorption profiles, as discussed in Section~\ref{s:obsTSAS-pulsar}.
Finally, several different approaches have been used to study spatial and temporal variability of optical and ultraviolet absorption lines and we summarize main results in Section~\ref{s:obsTSAS-optical}. 
We summarize key observational results from the HI absorption studies in Table 1 (non-pulsar sources) and Table 2 (pulsar sources) and list basic observed quantities of TSAS provided in the literature. 
We also provide in these tables the assumptions, such as distance to the absorbing medium and temperature, when available, with a goal of providing a complete data set
for future uses. For completeness, we list both detections and non-detections.
Several theoretical models and important questions about TSAS are outlined in Section~\ref{s:neutral_theory}.

\subsection{Spatial and temporal variability of HI absorption line
profiles against non-pulsar sources}
\label{s:obsTSAS-VLBI}

\subsubsection{Early Studies}
As irregularities in the ionized interstellar gas have been known to exist down to very small spatial scales, $\sim10^{7}$ cm, \citet{Dieter76} were the first to ask the question 
whether irregularities in the neutral medium could extend to similarly small scales.
Single, long baseline 
interferometric (VLBI) observations of the quasar 3C147 showed variations in the HI absorption line profiles with interferometric hour angle.
Dieter et al. (1976) interpreted these variations as being due to 
inhomogeneities of the absorbing medium. If resulting from a discrete cloud,
the measured difference in optical depth profiles $\Delta \tau$ implied a cloud HI column density of $N(HI)=10^{20}$ cm$^{-2}$ using:
\begin{equation}
N(HI)= C_0 \times T_s \times \int \Delta \tau dv
\end{equation}
where $C_0= 1.823 \times 10^{18}$ cm$^{-2}$ K$^{-1}$ (km/s)$^{-1}$) and $T_s$ is the excitation 
or spin temperature (the exact values for temperature and distance used in calculations are given in Table 1).
The velocity in this equation corresponds to radial velocity along the line of sight.
The maximum separation of source components of $0.16"$, 
implied a cloud size of $l=70$ AU and a HI volume density of $n_H=N(HI)/l=10^{5}$ cm$^{-3}$ (assuming spherical geometry).
The over-dense and over-pressured TSAS was born!
Subsequent VLBI observations by \citet{Diamond89}
found change in HI absorption spectra when varying angular resolution, implying
evidence for 25-AU HI clouds in directions toward 3C138, 3C380, and 3C147.
Particularly large variations were noticed in the case of
3C138, suggestive of TSAS with
$n_H=(5-10) \times 10^4$ cm$^{-3}$.
This study raised the questions regarding the filling factor, confinement, 
and regeneration mechanisms for such over-dense and over-pressured structures, 
and suggested that a continuous range of optically thick cloudlets 
with sizes of tens of AUs may exist in the ISM.

The first {\it{images}} of the HI optical depth distribution in the direction of
extragalactic sources were obtained by \citet{Davis96} toward 3C138 and 3C147, 
using the MERLIN array and the European VLBI Network, 
detecting significant optical depth fluctuations (Table 1).
\citet{Faison98} imaged 3C 138, 2255+416, and 0404+768 at even higher angular resolution 
of 10-20 mas using the Very Long Baseline Array (VLBA). They found optical depth variations for the first two sources. As  0404+768 did not show fluctuations over spatial 
scales probed (3--16 AU), they suggested a possible minimum size for TSAS structures of a few tens of AU.
\citet{Faison01} used the VLBA to extend  their previous study to seven sources in total.
They observed 3C147, 3C119, CJ1 2352+495 and CJ1 0831+557 with an angular resolution of 5 mas,
confirming only optical depth variation in the case of 3C147 (Table 1).
They also used Zeeman splitting to estimate the  upper limit on the (absolute) 
strength of the line-of-sight magnetic field of
32 $\mu$G for 3C147, 20 $\mu$G for 3C119, and 160  $\mu$G for 2352+495, suggesting that
the field is not enhanced towards these sources. 
However, considering a typical line-of-sight field of $\sim6$ $\mu$G \citep{Heiles05}, 
the obtained upper limits are clearly very high and further magnetic field measurements are highly needed.

In summary, while spatial variability of HI absorption profiles against extragalactic sources
found clear TSAS examples, since early days TSAS was not seen in all observed directions.
Faison et al. (1998) was the first to question whether the absence of TSAS on spatial scales below
few tens of AU has a physical meaning.  While most sources showed $\Delta \tau <0.1$,
3C138 and 3C147 showed much higher level of variability.

\subsubsection{The Highly Variable 3C138 and 3C147 Directions}

While several early observations showed significant optical depth variability
in the direction of 3C138 and 3C147, Brogan et al. (2005) and \citet{Lazio09} 
remain as studies with the most exquisite HI optical depth images against these sources.
\citet{Brogan05} imaged 3C138 in a three-epoch series of 
observations (1995, 1999 and 2002), and showed clear evidence for spatial variations of
typically $\Delta\tau\sim0.1$ (and reaching a maximum value of 0.5) on scales of 25 AU, Figure~\ref{f:3c138}.
In addition, {\it{temporal}} changes in the HI optical depth images have been found over a
period of 7 years, with implied transverse velocity for the intervening HI of order 20 \kms. 
In a followup study, Lazio et al. (2009) obtained VLBA HI absorption images of 3C147 and found
optical depth variations on typical scales of 10 AU (or 15 mas) with 
$\Delta \tau\sim0.1-0.3$ (reaching a maximum value of 0.7).

\begin{figure}
\includegraphics[scale=0.7]{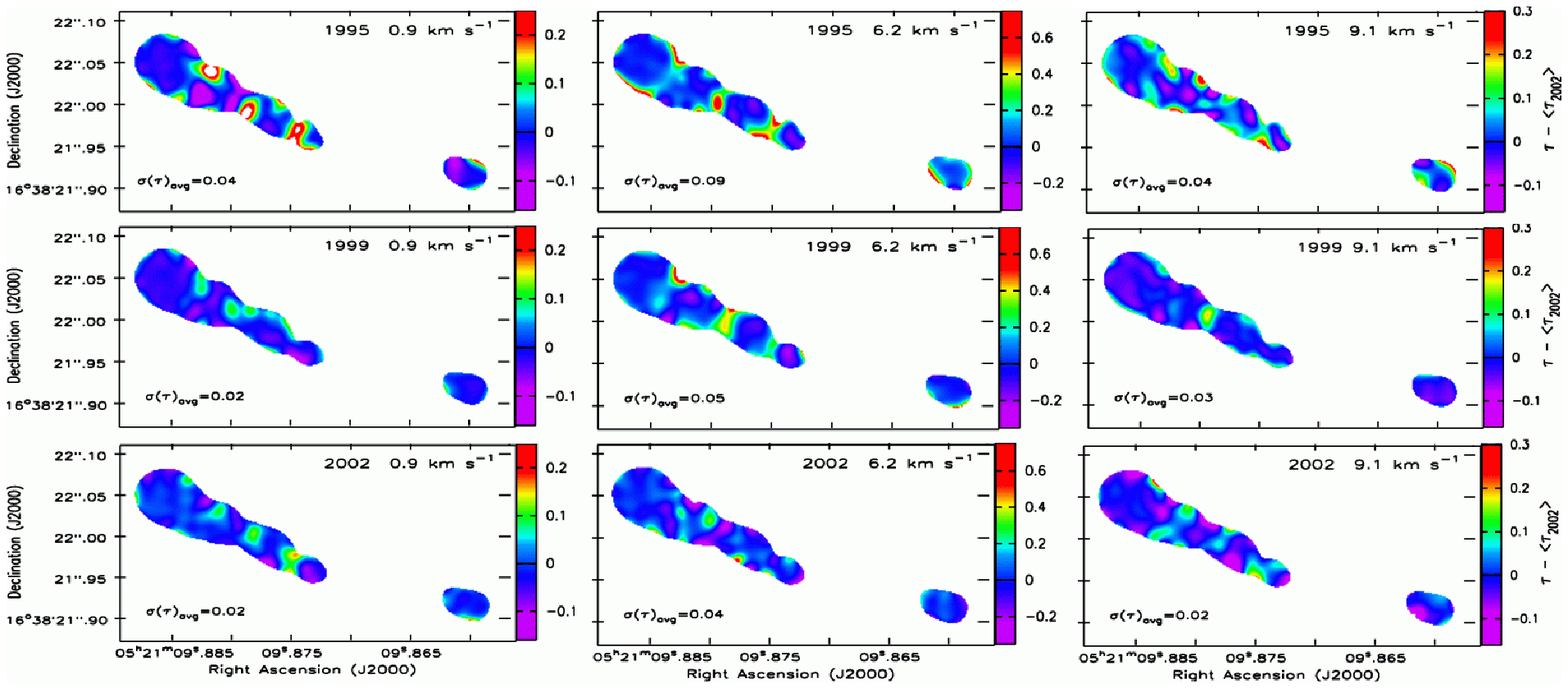}
\caption{The HI optical depth variability in the direction of 3C138 from Brogan et al. (2005). Images show the optical depth at three different velocities for three epochs. The the average value of the 2002 optical depth was subtracted from each image. The color scale has been adjusted so that the dark blue color is centered about zero. The average 
1-$\sigma$ uncertainty is indicated in the bottom left corner of each panel.}
\label{f:3c138}
\end{figure}

One possible way of mitigating the over-pressure problem of TSAS is if its 
spin temperature, $T_s$, is significantly lower relative to
the typical CNM clouds ($\sim15$ K, Heiles 1997). 
Brogan et al. (2005) compared line widths of HI absorption spectra 
between VLBA and 
\citet{Heiles03b} Arecibo observations of 3C138 and found excellent agreement. 
By assuming that linewidths are thermal in nature, they suggested that TSAS does not have significantly lower temperature than the typical CNM. 
However, we note that turbulent broadening can dominate over thermal broadening, resulting in the above experiment being inconclusive.  Therefore, 
the question of how TSAS compares to the CNM regarding its temperature  still remains as open.
For one of the velocity components in the direction of 3C138, 6.4 km/sec,
\citet{Heiles04} estimated
$B_{los}= +5.6 \pm 1.0$ $\mu$G. This is in agreement with the median total 
magnetic field (e.g. Heiles \& Troland 2005) 
supporting earlier suggestions that TSAS likely does not reside in regions with an enhanced
magnetic field strength and consistent with the theoretical arguments in \S~\ref{s:neutral_theory}.

Brogan et al. and Lazio et al. estimated the volume filling factor of TSAS in 
the direction of both 3C147 and 3C138 as being $\la1$\%.
This is in agreement with a few percent volume filling factor of the over-pressured 
ISM estimated by Jenkins \& Tripp (2001), and such structures are expected to be 
far from equilibrium and short-lived. However, according to the
energy input argument given in \S 1.2, the energy input from free expansion of such structures
would be of order $2.0\times 10^{-22}(P_c/10^{-11})(v_5)(10 ~AU/l)$ erg cm$^{-3}$s$^{-1}$, where the units of pressure and velocity are in dyne cm$^{-2}$ and km s$^{-1}$.  Although this is an upper limit, the volume filling factor of 1\% implies that this heating rate can only be balanced by
radiative cooling for $n > 370/T_2^{0.4}$ cm$^{-3}$, which is very high for the CNM.
This large heating rate would affect many areas of astrophysics, and would have a detectable effect on temperature of dust grains and HI clouds.


While both studies claimed that sightlines to 3C138 and 3C147 are not special,
previous HI emission observations of 3C147 by \citet{Kalberla85} found HI at 500-2000 K -- this thermally
unstable HI likely signals possible dynamical events in the recent history of the region.
In addition, we note that both directions have significant HI column density:
$\sim 6 \times 10^{21}$ cm$^{-2}$ for 3C138 
and $\sim 8 \times 10^{20}$ cm$^{-2}$ for 3C147.
This is higher than the column density of typical diffuse lines of sight and 
is similar to what is typically found close to molecular clouds \citep{Stanimirovic14}.
Such column densities may have enough shielding for H$_2$ formation, e.g. \citet{Lee12},
and could signal the presence of a denser, more clumpy medium.

Brogan et al. suggested that TSAS is ubiquitous in the ISM. 
Their explanation for the lower observed  levels of variations 
in the  cases of sources other than 3C138 and 3C147 is based on  selection 
effects, e.g. other sources have a smaller angular extent and lower surface brightness,
resulting in a smaller dynamic range of probed spatial scales.
However, the typical size of TSAS structures found for both 3C138 and 3C147 is only 2--2.5 times 
larger than the telescope beam (although observations were sensitive 
to a continuous range of spatial scales from 20 to 300 mas).
Future observations with a broader observed range of TSAS sizes are needed to confirm this interpretation.

\begin{table}
\caption{Summary of TSAS radio detections and non-detections for non-pulsar sources.}
\label{t:TSAS}
\begin{center}
\begin{tabular}{@{}l|c|c|c|c|c|c@{}}
\hline
Source & Coords ($^{\circ}$) & Size (AU)& $\Delta \tau$ & $N$ (cm$^{-2}$)& Comment\\
\hline
3c147$^a$     &161.69,10.30  &  70   & -             & $10^{20}$ & $d=600$ pc, $T_s=50$ K  \\\hline
several$^a$   &- &  -    &               & &           no detection    \\\hline   
3c138$^b$     &187.5,$-11.3$  &  25   &               & $10^{19.6}$    &  $d=500$ pc, $T_s=50$ K   \\\hline
3c147$^b$     &161.69,10.30  &  25   &  some         & &                   \\\hline
3c380$^b$     &77.23, 23.50  &  25   & some          & &                   \\\hline
3c138$^c$     &187.5,$-11.3$  &  -    & 0.16,0.35     & &                 \\\hline
3c147$^c$     &161.69,10.30  &  -    & 0.07          & &                   \\\hline
3c138$^d$     &187.5,$-11.3$  &       & 0.6-0.8          & &                \\\hline
2255+416$^d$  &101.23,$-16.11$  &       & some          & &                  \\\hline
0404+768$^d$  &133.41,18.33  & 3-16  &               & &      no detection  \\\hline
3c147$^e$     &161.69,10.30  &       & 0.28,0.2      &          &  $d=1,0.5$ kpc   \\\hline  
3c119$^e$     &160.96,$-4.34$  &10-100 &$<0.2$         &          &  $d=1.3$ kpc     \\\hline
2352+495$^e$  &113.71,$-12.03$  &5-40   &$<0.1$         &          &                   \\\hline
0831+557$^e$  &162.23,36.56  &30     &$<0.1$         &          &                \\\hline
3C138$^f$&187.5, $-11.3$& 25& $0.1\pm0.05$&$10^{19.5}$    &$d=500$ pc, $v_{ch}=0.82$ \\\hline
3C147$^g$&161.69, 10.30 &10 & $0.2\pm0.07$&$10^{19.7}$    &$d=750$ pc, $v_{ch}=0.4$  \\\hline
3C161$^h$&215.44,$-8.07$&450&$1.25\pm0.25$&$10^{20.5}$    &$d=1.8$ kpc,$v_{ch}=0.49$ \\\hline
3C123$^h$&170.58,$-11.66$&107&$0.7\pm0.25$& $10^{20.3}$   &$d=530$ pc, $v_{ch}=0.49$\\\hline
3C111$^h$&161.68,$-8.82$&42000&0.3	&      $10^{19.9}$   &		\\\hline
GRS1915$^i$&45.37,$-0.22$&900&0.67	&	            &55 km/s, $d=6.1$ kpc, $v_{ch}=2.6$\\\hline
GRS1915$^i$&45.37,$-0.22$&440 &0.24	&	            &5 km/s, $d=11$ kpc, $v_{ch}=2.6$\\\hline
GRS1915$^i$&45.37,$-0.22$&150-900&$<0.5$&	     &55 km/s, $d=6.1$ kpc, $v_{ch}=2.6$\\\hline
GRS1915$^i$&45.37,$-0.22$&275-1650&$<0.25$&      &5 km/s, $d=11$ kpc, $v_{ch}=2.6$\\\hline
3C405$^j$&76.2,5.7&	  206265. &0.13 &	     &	$d=1.1$ kpc, $v_{ch}=2.75$\\\hline
six other$^j$&    &	6000-77000&$<0.03-0.2$&	     &		no detections\\\hline
3C018$^k$&118.62,$-52.73$&6077&$0.095\pm0.007$&	   &    $v_{ch}=0.4$ \\\hline
3C041$^k$&131.38,$-29.07$&4772&$0.036\pm0.007$&	   &	$v_{ch}=0.4$\\\hline
3C111$^k$&161.68,$-8.82$& 78649&$0.377\pm0.008$&	   &	$v_{ch}=0.4$\\\hline
3C123$^k$&170.58,$-11.66$&10631&$0.105\pm0.0035$&	   &    $v_{ch}=0.4$\\\hline
3C225$^k$&220.01,44.01&692&$0.040\pm0.004$&	   &    $v_{ch}=0.4$\\\hline
3C245$^k$&233.12,56.30&530&$<0.02$&		       &	$v_{ch}=0.4$\\\hline
3C327.1$^k$&12.18,37.01&1946&$0.093\pm0.012$&	   &    $v_{ch}=0.4$\\\hline
3C409$^k$&63.40,$-6.12$&5516&$0.275\pm0.007$&	   &	$v_{ch}=0.4$	\\\hline
3C410$^k$&69.21,$-3.77$&8155&$0.523\pm0.013$&	   &	$v_{ch}=0.4$\\\hline

\hline
\end{tabular}
\end{center}
\begin{tabnote}
Column 2 - source Galactic coordinates in degrees. Column 3 - spacial scale of TSAS in AU.
Column 4 -variation in HI optical depth. Column 5 - HI column density in cm$^{-2}$. 
Column 6 - comments include assumed TSAS distance and spin temperature, when provided. 
The velocity resolution is listed as $v_{ch}$ as it is important to place all measurements on the same scale.
References: $^a$ - \cite{Dieter76}; $^b$ - \cite{Diamond89}; $^c$ - \cite{Davis96}; $^d$ - \cite{Faison98};
$^e$ - \cite{Faison01}; $^f$ - \cite{Brogan05};  
$^g$ - \cite{Lazio09}; $^h$ - \cite{Goss08};
$^i$ - \cite{Dhawan00}; $^j$ - \cite{Dickey79}; $^k$ - Murray et al. (2015)
\end{tabnote}
\end{table}

\begin{table}
\caption{Summary of TSAS radio detections and non-detections for pulsars.}
\label{t:TSAS2}
\begin{center}
\begin{tabular}{@{}l|c|c|c|c|c|c@{}}
\hline
Source & Coords ($^{\circ}$) & Size (AU)& $\Delta \tau$ & $N$ (cm$^{-2}$)& Comment \\
\hline
PSR B1821+05$^a$&34.99,8.86&75	  &$2\pm0.85$&	     &		$v_{ch}=1.2$ \\\hline
PSR B1557-50$^b$&330.7,1.6 &1000 &1.1	     &	     &		$v_{ch}=6.7$	\\\hline
PSR B1154-62$^b$&296.71,$-0.20$          &	&		     &    & no detection\\\hline
PSR B1557-50$^c$&330.7,1.6&1000	&$0.15\pm0.05$&	     &		$d=6.4$ kpc, $v_{ch}=7.3$ \\\hline
three other PSRs$^c$&      &50-100&$<0.02-0.12$&      &  no detection\\\hline
six PSRs$^d$&       &5-100&	0.03-0.7&  $10^{19-20.7}$          & $T_s=50$ K \\\hline
PSR B0301+19$^e$&49.20,2.10&500	&$0.15\pm0.01$&	     &          $v_{ch}=2$\\\hline
two PSRs$^e$&        &	&$<0.018-0.14$&      &	no detection \\\hline
PSR B0329+54$^f$& 144.99,$-1.22$ &0.005-25 & $<0.026$      &    & no detection \\\hline
PSR B1929+10$^g$&47.38,$-3.88$&6.&$0.150\pm0.008$& $10^{19.1}$& $T_s=170$ K, $v_{ch}=1.04$\\\hline
PSR B1929+10$^g$&47.38,$-3.88$&12.&$0.025\pm0.008$&$10^{19.4}$& $T_s=170$ K, $v_{ch}=1.04$\\\hline
PSR B1929+10$^g$&47.38,$-3.88$& 28.&$0.020\pm0.008$&$10^{19.1}$& $T_s=170$ K, $v_{ch}=1.04$\\\hline
PSR B1929+10$^g$&47.38,$-3.88$& 46.&$0.020\pm0.008$&$10^{19.4}$& $T_s=170$ K, $v_{ch}=1.04$\\\hline
PSR B2016+28$^g$  &68.10,$-3.98$  & 1-10  & $<0.1-0.2$ &   &  $v_{ch}=1.04$   \\\hline
PSR B0823+23$^g$  &196.96,31.74  & 5-50  & $<0.01-0.02$& &   $v_{ch}=1.04$   \\\hline
PSR B1133+16$^g$  &241.90,69.19  & 20-170& $<0.01-0.02$& &  $v_{ch}=1.04$  \\\hline
PSR B1737+13$^g$  &37.08,21.68  & 20-180& $<0.1-0.2$&   &   $v_{ch}=1.04$   \\\hline

\hline
\end{tabular}
\end{center}
\begin{tabnote}
Same as in Table 1.
References: $^a$ - \cite{Clifton88};
$^b$ - \cite{Deshpande92}; $^c$ - \cite{Johnston03}; $^d$ - \cite{Frail94}; $^e$ - \cite{Weisberg08}; $^f$ - \cite{Minter05}; $^g$ - \cite{Stanimirovic10}
\end{tabnote}
\end{table}

In summary, 3C138 and 3C147 have shown significant spatial and temporal variations of HI optical depth over several studies.   
TSAS features in these two directions show interesting similarities.
TSAS volume filling factor  is $\sim1$\%, however no clear observational constraints of the TSAS temperature exist to date.
We caution that in both cases, the typical TSAS size found is very close to the linear resolution of VLBA observations. Both directions have the total HI column density 
similar to what is found close to molecular clouds,
suggesting suitable conditions for the formation of H$_2$. 
As many numerical simulations suggest H$_2$ formation via converging 
flows \citep[e.g.][]{Hennebelle07}, both lines of sight could be undergoing a post-shock phase transformation
which is characterized by out-of-equilibrium physical properties. 
If this is the case, then such a level of optical depth variations may not be typical for the ISM. 
Future observations should search for velocity signatures of converging flows or other dynamical imprints on the velocity spectra.

\subsubsection{Additional Sources}

\begin{figure}
\includegraphics[scale=0.6]{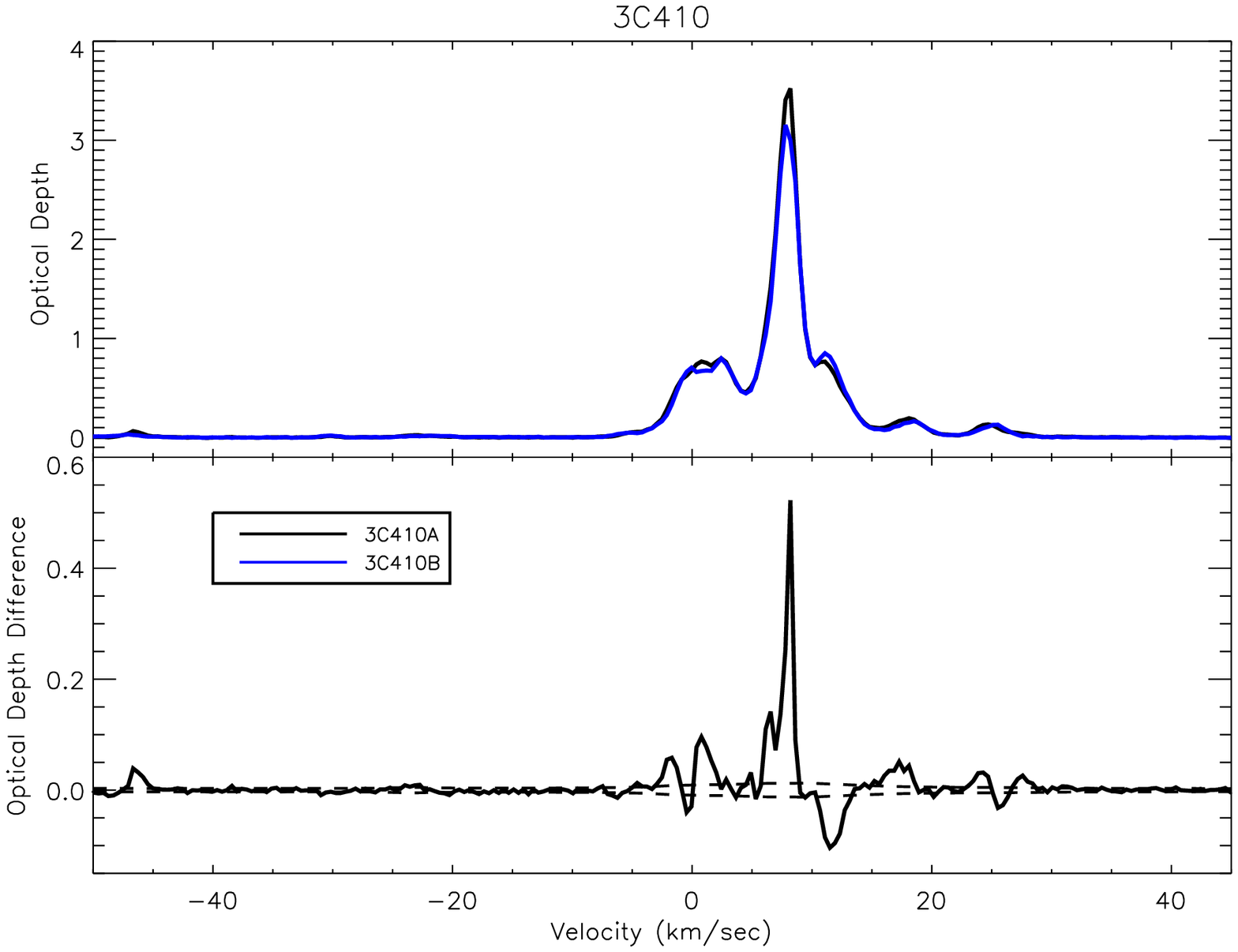}
\caption{(Top) HI absorption spectra in the direction of radio continuum sources 3C410A and 3C410B
from the 21-SPONGE survey shown in black and blue.  The two components (A and B) in this double source
are separated by only 5.4$"$. At the assumed distance of $0.1/\sin b=1.5$ kpc, 
the separation between two components
corresponds to a linear distance of $\sim8,000$ AU.  
(Bottom) The difference of two HI optical depth profiles. Dashed lines show the 1-$\sigma$ uncertainty.
The maximum $\Delta \tau=0.5$ is found at a radial velocity of $\sim 8$ km/s. 
}
\label{f:3c410}
\end{figure}

\citet{Dhawan00} used an exotic source - a microquasar GRS 1915+105 with a  proper motion of 8-17 mas day$^{-1}$ - to look for temporal changes in HI absorption spectra. At two different radial velocities, they found significant optical depth variations (for details see Table 1).
\citet{Goss08} used MERLIN to obtain interferometric imaging of  three sources: 3C111, 3C123 and 3C161. In the case of 3C161 and 3C111 they found significant optical depth variations 
on spatial scales of $\sim400-500$ AU and $\sim42,000$ AU, respectively.
3C111 is especially interesting as the same source showed
variability in H$_2$CO absorption by \citet{Moore95}. 
On the other hand, 3C123 showed only hints of variability at a low statistical significance.

To probe even larger spatial scales of $>10,000$ AU (often studied with optical absorption against stars in clusters), observations of double-component extragalactic radio sources can be used. \citet{Dickey79} obtained interferometric HI absorption observations
in the direction of 9 double sources and found a significant difference in 3 sources with one source (3C405)
being especially convincing with $\Delta \tau \sim 0.13$ at an upper 
limit for the linear scale of HI optical depth fluctuations of $\sim1$ pc\footnote{This study and many others derived the upper limit on the distance of the absorbing TSAS as: 110 pc/sin$|b|$, by assuming the CNM scale-height of 110 pc.}. This study, however, concluded that small-scale HI optical depth structures
 are uncommon, supporting
the validity of 21-cm spin temperature measurements using beam switching.

Very recently, 21-SPONGE has obtained HI absorption spectra in the direction
of 52 radio continuum sources with the VLA at exceptional sensitivity,
$\sigma_{\tau}<0.001$ per 0.4 km/s channels \citep{Murray15}.
Out of 52 sources 8 sources are doubles and one is a triple.
This provides a unique opportunity to look for HI optical depth differences
at close separations. Of these nine sources 
eight show significant differences on spatial scales from 700 to 79,000 AU,
with    $\Delta \tau$ ranging from 0.04 to 0.5 (Stanimirovic et al., in prep) and 
a tendency for larger variations to be found at larger spatial separations 
As an example, we show the case of 3C410A/B in Figure~\ref{f:3c410}.
We note excellent agreement for 3C111 between this study and Goss et al. (2008), while
3C123 observed by Goss et al. had a higher $\Delta \tau$. 

A new type of HI structures, potentially related to larger TSAS, was discovered recently 
by \cite{Clark14} by applying the Rolling Hough Transform on the HI emission observations. 
These linear HI features, named ``HI fibers'', likely trace the CNM
at spatial scales $<10^4$ AU, $N(HI)\sim5 \times 10^{18}$, and are oriented along the magnetic field lines.
\cite{Clark14} suggest that HI fibers are likely associated with the Local Bubble wall.

Finally, we note that \citet{Dieter-Conklin09} emphasized the idea by \citet{Marscher93} that due to the 
Earth's motion around the Sun, the Solar motion in the Galaxy, 
and the cloud's proper motion, our line of sight to a distant background
source is continuously (slowly) moving and we are sampling an intervening ISM at different locations.
While this line-of-sight (or ``searchlight'') wandering  still samples TSAS, the main point is that
observations are tracing varying sightlines in an interstellar cloud, as opposed to the same 
sightine where physical properties
of the cloud have changed over time. This agrees with the Deshpande (2000) idea
that observations at different epochs are sampling a
point on the structure function not the power spectrum of optical depth fluctuations.

\subsubsection{Power Spectrum of the HI Optical Depth}

While most reported studies used discrete (or limited) measurements of the HI optical depth 
and did not have high enough spatial dynamic range to
calculate the power spectrum, several direct measurements
of the $\tau$ power spectrum exist to date. 
Using the SNR Cassiopeia A (or Cas A) as an extended background source, \citet{Deshpande00b}
measured the spatial power spectrum of the HI optical depth images 
($\tau(v)$, as well as $\int \tau(v)~dv$) over several different velocity ranges and
found a slope of $2.75\pm0.25$  over a range of spatial scales 
from 0.07 to 3 pc\footnote{A distance of 
2 kpc was assumed for the Perseus arm.}.
For Cygnus A they found similar slope for the Outer Arm, while an even more shallow slope of 2.5
for the Local Arm.
This power-law index for Cassiopeia A, obtained with the VLA, was later confirmed in an 
independent experiment by \citet{Roy10} using the Giant Metrewave Radio Telescope.
Several observational studies have claimed the observed $\Delta \tau$, or upper limits, in agreement with
the Cas A optical depth spectrum when extrapolated down to $\sim100$ AU \citep{Dhawan00,Faison01,Johnston03}.

\citet{Roy12} combined VLBA, Merlin and VLA interferometric observations of 3C138 to probe a range of angular scales
from 10 mas to $0.2^{''}$\footnote{They used a distance of 500 pc which results in the range 
spatial scales of 5-100 AU.}. They calculated a structure function of 
the HI optical depth images, $\tau(v)$, for individual velocity channels
and found a corresponding power-spectrum slope of $2.33\pm0.07$.
No significant channel-to-channel variations of the structure-function slope were found.
\citet{Dutta14} used the same data set  but with Monte Carlo simulations
to assess the effect of low S/N data on the structure function slope.
They found a slightly steeper slope, $2.81^{+0.14}_{-0.13}$, in excellent agreement with
the Deshpande et al. (2000b) Cas A results on much larger scales. 

Combining results for these three different 
regions (Cas A, Cygnus A  and 3C138) hint at a possibly similar power spectrum
slope of the HI optical depth over a range from 
5 AU to 3 pc, which is impressively six orders of magnitude! 
This result clearly needs to be confirmed and additional observations obtained to bridge the 
current gap that exists on spatial scales $\sim10^2$ to $\sim10^4$ AU. 
Extrapolation of the HI optical depth power spectrum to very small scales has been 
one of the key criticisms of the Deshpande (2000) model - the measured structure function 
slope in the direction of 3C138 at least somewhat resolves this concern.
However, the question of whether the same slope applies for the cold HI throughout the 
whole Milky Way still remains.
As Brogan et al. (2005) noticed, a small change in the slope of just 0.1 implies a 
change in the optical depth variations of a
factor of 2. Therefore testing the uniformity of the structure function slope 
with future measurements is very important.

\cite{Deshpande00b} argue that the slopes of the power spectra of relative density fluctuations $\delta n_H/n_H$ and optical depth
variations $\Delta\tau/\tau$ are the same (this argument assumes both are small). 
The fraction of mass $\delta M(<l)/\delta M$ in
fluctuations of size $< l$ for any scale $l$
 within the spectrum is related to the power law index $q$ by $\delta M(<l)/\delta M\sim (l/l_{max})^{q+2}=(l/l_{max})^{4.75}$. That is, the mass in fluctuations is dominated by the largest fluctuations. 


\subsection{Temporal  variability of HI absorption profiles against pulsars}
\label{s:obsTSAS-pulsar}

There are several reasons why pulsars have been recognized as ideal sources for studying TSAS.
First, pulsars have
a relatively high proper motion and transverse speeds (typically $\sim10-100$ AU yr$^{-1}$),
sampling the CNM on AU spatial scales by obtaining absorption profiles at different epochs. 
Second, the pulsed nature of pulsars' emission allows spectra to be obtained {\it on} and {\it off} source
without moving the telescope, therefore sampling both emission and absorption along 
almost exactly the same line of sight \citep{Stanimirovic10}. With radiative transfer,
the on- and off-source spectra can then be used to estimate $T_s$. 
Third, the compact nature of pulsars  samples the gas with an effective absorption
beam that is limited only by interstellar scattering (see Section~\ref{s:TSIS-general}). 
This typically gives resolution of 0.1-1 mas \citep{Dickey81}. 
However, the compact nature of pulsars, as well as their variable nature,  
require   sophisticated fast-sampling 
spectrometers and careful data processing.
For example, interstellar scintillation can cause 
baseline ripples, while the varying source nature can lead to
``ghost effects'' \citep[artificial absorption features,][]{Weisberg80}, which need to be removed.

A unique and exceptionally important aspect of pulsars is that they can be used to probe neutral,
ionized, and  molecular  medium along {\it identical} lines of sight. This is especially interesting as high-density TSAS may contain molecules, 
and/or have ionized outer layers that could be observed as dispersion measure (DM) variations. 
While HI has been studied in absorption using pulsars for several decades, recent
studies have detected OH in absorption against several pulsars \citep{Stanimirovic03a,Weisberg05}.
Only two studies have so far obtained simultaneous measurements of HI absorption and dispersion measure 
\citep{Frail94}. This is clearly an open area where important progress can be made in the future.

\subsubsection{Early Studies}
In the late 1980s, sufficiently accurate and repeated measurements of HI absorption profiles
against pulsars began to be made, and it was noticed that pulsar ISM spectra changed over time in some cases,
suggesting inhomogeneities in the intervening gas.  For
example, \citet{Clifton88} found that the HI absorption spectrum of PSR B1821+05
changed between $\sim 1981$ and 1988 by $\Delta \tau \sim2$. 
\citet{Frail91} noticed large optical depth variations towards the same pulsar.
\citet{Deshpande92} showed that between $\sim 1976$ and 1981, HI
absorption toward  B1154-62 did  not change significantly; while toward
B1557-50, a variation with $\Delta\tau\sim 1.1^{+1.1}_{-0.6} $ was interpreted as a TSAS of
size in the 1000 AU range.

The early pulsar HI results inspired Frail et al. (1994)
to undertake the first dedicated multi-epoch pulsar HI absorption experiment at
Arecibo.  Six pulsars were observed at three epochs, probing spatial scales 5--100 AU.  
These authors reported the presence of {\it{pervasive}} variations with $\Delta\tau\sim0.03-0.7$  
for all six pulsars.  They indicated that TSAS could comprise 10-15\% of the cold ISM.
For one of the Frail et al. (1994) pulsars, B0823+26, DM variations of $+0.0013$ pc cm$^{-3}$ were measured 
by \citet{Phillips91} at the same time as HI absorption. DM variations 
correspond to $\delta N_{H^+} =4 \times 10^{15}$ cm$^{-2}$, the level much smaller than
the observed variation in HI column density of $\delta N_{HI} =-5.5 \times 10^{18}$ cm$^{-2}$.
Their conclusion was that DM variations are smooth in this direction and not
driven by large-scale fluctuations, while TSAS was significant. 

While more recent studies did not confirm TSAS abundance and properties claimed by Frail et al.,
and have questioned spectrometer accuracy (Stanimirovic et al. 2010) and statistical significance of some of their detections (Johnston et al. 2003), the Frail et al. study provided an important impetus for future experimental and theoretical work.

\subsubsection{More Recent Studies}
The recent era of  sensitive pulsar TSAS experiments began with the Parkes observations
of  \citet{Johnston03}.  These investigators found no significant
optical depth variations in their three-epoch
observations of three pulsars (0736-40, 1451-68 and 1641-45, for limits see Table 1).
In comparing their work with previous studies, they 
showed that Frail et al. (1994) did not fully account for the large increase in noise 
in absorption spectra at the line frequency, suggesting that some of Frail et al. detections were not real.
They found TSAS only in the case of one pulsar: PSR B1557-50. This
is the same pulsar whose HI spectral variability was noted earlier by Deshpande et al. (1992).
In combining the results from four measurements over twenty-five years, 
Johnston et al. (2003) found a TSAS feature $\sim 1000$ AU in size, 
with $n_H \sim 10^4$ cm$^{-3}$.
They explained their detections and non-detections as agreeing with the Deshpande (2000) picture.

\begin{figure}
\includegraphics[scale=0.5]{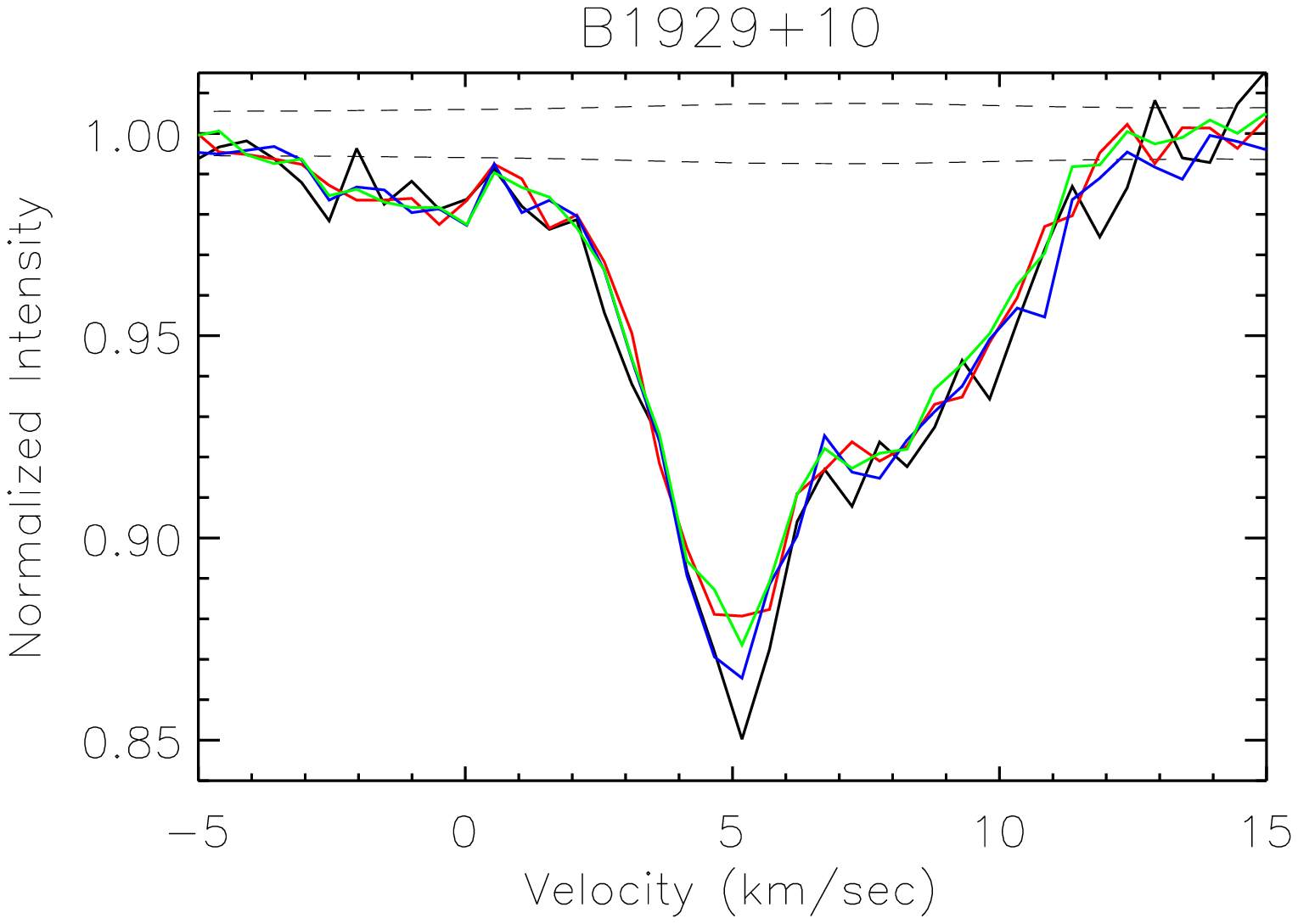}
\caption{HI absorption spectra in the direction of B1929+10 obtained with the
  Arecibo telescope at four observing epochs. Dashed lines show a typical, $\pm1-\sigma$ noise
  level in the absorption spectra. The absorption feature at 5 \kms has
  changed significantly (2-3-$\sigma$ level) with time \citep{Stanimirovic09,Stanimirovic10}.}
\label{f:1929}
\end{figure}

\citet{Minter05} performed a comprehensive TSAS search in the direction of PSR B0329+54 with the
Green Bank Telescope.
The pulsar was observed in eighteen sessions over a period of 1.3 yr, yet
no HI optical depth variations were detected 
for pulsar transverse offsets ranging from 0.005--25 AU\footnote{For consistency 
we normalized scales to assume that absorbing material is at the position
of the pulsar which is a common assumption for many studies.}. 
This study also tested the \citet{Gwinn01} explanation of small-scale structure as being
due to a combination of interstellar scintillation and gradients in the Doppler velocity of HI 
but did not find support for this model. 
While no HI optical depth variations were observed, \citet{Shishov03}
have obtained diffractive scintillation measurements for this pulsar.
They found that below 2000 AU the diffractive scintillation can be explained
with a scattering screen comprised solely of ionized gas.
On scales larger than 2000 AU their results require some neutral gas
to be present inside the scattering screen. Minter et al. suggested that therefore the inner
scale of neutral gas could be around 2000 AU and that this could explain the lack of
HI absorption variability below 2000 AU.

In their HI absorption study of pulsars in the first Galactic quadrant,  \citet{Weisberg08}
noticed that one of three pulsars with previous HI absorption measurements,
PSR B0301+19, exhibited significant changes in its absorption 
spectrum over a period of 22 yr, indicating TSAS on a 500 AU scale. 

\subsubsection{Variability in the direction of B1929+10 and the ubiquity of TSAS}
A new multi-epoch pulsar experiment re-observed Frail et al. (1994) targets
at even higher sensitivity \citep{Stanimirovic03b,Weisberg07,Stanimirovic10}. 
B0540+23 was excluded from the analysis due to strong interstellar scintillation causing complex baseline ripples. TSAS was detected only in the case of B1929+10 and only for
the strongest of its three velocity components  (shown in Figure~\ref{f:1929}),
when considering both equivalent width of the HI optical depth and 
difference spectra between two epochs. 
Table 2 lists information for all pulsar detections and non-detections.
Four detected TSAS features have a size of
6--40 AU, $N(HI)=(1-3) \times 10^{19}$ cm$^{-2}$, $n_H>10^4$ cm$^{-3}$ 
(assuming spherical geometry) and $P/k> {\rm a ~few} \times 10^6$ cm$^{-3}$ K.
The fraction of HI associated with TSAS was found to range from 0\% to 11\%, 
with a median value of 4\%.
In addition, by combining their measurements and upper limits with published results, 
\cite{Stanimirovic10} found that the maximum detected $\Delta \tau$ showed an increasing trend 
with the peak optical depth. 

Stanimirovic et al. (2010) found that derived $T_s=150-200$ K for HI components in the direction of B1929+10
are significantly warmer than what is typically found for the CNM (e.g. 20-70 K). In addition, this
line of sight has only 7\% CNM fraction of the total neutral gas, while other sightlines have 15-20\%.
B1929+10 has a distance of only $361^{+8}_{-10}$ pc and is the closest in the observed sample with
one half of its line of sight being inside the Local Bubble, running along the
Local Bubble wall for $\sim 50-60$ pc.
Based on a comparison with Na I observations, it is likely that the absorbing clouds traced via 
varying optical depth are at a distance of
$<106$ pc \citep{Genova97}, 
confirming their association with the Local Bubble, in support of the measured high $T_s$.

The examination of all pulsar and interferometric TSAS detections and upper limits
by \cite{Stanimirovic10} did not show a correlation between the level of optical depth fluctuations and the TSAS spatial scale, as would be expected if the turbulent spectrum on
much larger scales is extrapolated to AU-scales.
The detections and non-detections probed an almost continuous range
of spatial scales from $\sim0.1$ to 1000 AU.
This study concluded that the large number of non-detections of TSAS suggests that the CNM clouds 
on scales $10^{-1}$ to $10^{3}$ AU are not a pervasive property of the ISM.
The sporadic TSAS detections on scales of tens of AU could indicate an intermittent
process that forms discrete structures or turbulence associated with large scale features (such as 
the Local Bubble Wall) rather than end-points of a universal turbulent spectrum. 

Another striking result from their study was a possible correlation between B1929+10's TSAS and
 interstellar clouds observed in Na I absorption inside the Local Bubble.
There is significant  evidence that the TSAS in the direction of B1929+10 
is likely to be within $\sim100$ pc of the Sun, and is
sampling the small-scale structure of the Local Bubble likely caused 
by hydrodynamic instabilities fragmenting the Local Bubble wall.
Stanimirovic et al. (2010)  proposed that the line of sight of B1929+10 is revealing
this recently formed TSAS. Similar bubbles and their walls are found throughout
 the Milky Way,  but the lifetime of a TSAS cloud created from them depends strongly on the 
 its size and the temperature of the surrounding medium.
Larger fragments (size $\sim10^4$ AU) survive longer and can travel large ISM distances, 
becoming a more general ISM property. On the other hand, the smallest clouds (size $\sim10-100$ AU)
evaporate quickly close to their formation site, and are therefore
not very commonly observed in the ISM. Additional processes, such as stellar
mass-loss and collisions
of interstellar clouds/filaments, could also contribute to
the CNM structure formation on somewhat larger (sub-pc) scales.

In summary, the preponderance of evidence in recent years (and increased sensitivity of radio instruments) suggests that TSAS is {\textit{not}} ubiquitous. What physical and environmental conditions, and spatial scales are more conducive for TSAS formation and evolution is still not clear. While several pulsar measurements detected TSAS on spatial scales of $\sim10$ AU, in one direction (B0329+54) it was suggested that the inner scale of TSAS is more likely to be 2000 AU. There is significant evidence that TSAS in the direction of B1929+10 is within the Local Bubble and likely associated with the fragmenting Local Bubble wall.

\subsection{Spatial and temporal variability of optical and ultraviolet lines}
\label{s:obsTSAS-optical}

Several different methods have been employed to study TSAS using optical and ultraviolet
transitions. Spatial variability of line profiles
in the direction of stars in binary and multiple stellar systems typically probes
spatial scales of a few thousands of AU. With modern multi-object
spectrographs, even 2D images of the absorbing medium can be reproduced using
hundreds of stars in globular and open clusters. In addition, temporal variability
of absorption lines against stars probes even smaller spatial scales of $\sim 1$ to tens of AUs, Figure 1. 

The most commonly used transitions for such studies are Na I lines at 5889 
and 5895 \AA~ due to their strengths
and availability from the ground, relatively high gas phase abundance, as well as the wavelengths being at the
peak of efficiency of high-resolution spectrographs \citep{Lauroesch07}.
Other commonly used species are Ca I, Ca II,  K I, and even molecular lines like CH and CN.
In more recent years, several studies have observed the 
diffuse interstellar bands (DIBs) -- over 400 broad optical absorption features in the 4000-9000 \AA~ range whose 
origin is still debated but are likely produced by carbonaceous molecules.
In addition, many UV lines, from dominant ions such as Mg I, Cr II, Zn II, S II, are available using space-based observations and can provide useful diagnostics via line ratios.

The key difference between radio and optical observations of TSAS is that
lines in the optical come mainly from trace ionization states (Na I, K I) and elements
which are highly depleted onto dust grains even in diffuse clouds.
This makes the interpretation of optical variability very  complex as
line profile variations can reflect changes in the column density, but also variations
in the physical conditions (such as density, ionization history, temperature, or even  chemistry).
In our summary of main results from optical/UV observations of TSAS we follow somewhat an excellent review provided by Lauroesch (2007).

\subsubsection{Early Studies}

The very first studies to note variations in the optical line profiles  over small 
angular scales go back all the way to \citet{Munch53} and \citet{Munch57}.
The binary stars observations by \citet{Meyer90,Meyer94}
found  variability of Ca II absorption lines with a higher frequency on 
scales larger than $2500$ AU.
Using the Anglo-Australian Telescope \citet{Meyer96} observed five binary pairs and found Na I 
variability in all cases on spatial scales of 2800--12,300 AU.
Especially striking were two stars in $\mu$ Cru, separated by 6600 
AU\footnote{A distance of 170 pc was used in this study.},
where the component at 5 km/s has changed in column density by $2 \times 10^{11}$ cm$^{-2}$.
Although Na I is not a dominant ion in HI clouds, an empirical relation exists  between $N(Na I)$ and $N(H)$.
This relation implied a change in the H column density of $N(H)>10^{20}$ cm$^{-2}$ and the
TSAS number density of  $n_H>10^3$ cm$^{-3}$.
It has been known, however, that the $\mu$ Cru sightline passes through an expending shell 
associated with Loop I.  Building further statistics, \citet{Watson96} studied 15 binary and 
two triple systems covering projected separations of 480 to 29,000 AU and found significant variations for all  systems. However, they also found several cases where variability was occurring for some, but not all,  velocity components. Similarly to early radio observations of TSAS, they concluded that TSAS traced by 
Na I is likely ubiquitous.

\citet{Lauroesch03} observed temporal variability over a period of 8 years
toward $\rho$ Leo using Na I and Ca II transitions. 
They found that the  velocity component at 18 km/s showed significant temporal variability in Na I on scales of $\sim12$ AU, however Ca II stayed unchanged.
This prompted them to use archival Hubble Space Telescope (HST) observations of other species 
to measure thermal pressure of  $P/k<10^{3.3-3.8}$ cm$^{-3}$ K 
and $n_H<20$ cm$^{-3}$ using C I lines. These results, both for pressure and density,  
are significantly lower than what is traditionally assumed for the overpressured models of TSAS. 
Furthermore, the Cr II to Zn II ratio and the estimated
electron density all suggested properties typical for diffuse clouds.
This study clearly highlighted complications when dealing with Na I -
the Na I column density changes in this particular direction were 
clearly driven by changes in physical conditions
of the absorbing gas, not the total gas column density, 
which would happen in the presence of discrete structures. 
They suggested a picture where the bulk of trace species are in density peaks so their
patchy distribution gives an appearance of large fluctuations on AU scales.
Similar results were obtained in other studies. 
For example, \citet{Lauroesch98} used HST observations 
to show that for $\mu$ Cru common neutral species show variations, while dominant ions do not. 
Further studies by \citet{Lauroesch99,Pan01,Welty07} especially argued for density variations and/or ionization variations
within larger interstellar clouds as being the cause of observed column density differences.

More recent studies have employed even higher spectral resolution multi-object and integral-field 
spectrographs (e.g. WIYN's DensePak) to map Na I absorption in the direction of 
many stars, providing essentially 2-D images of the absorbing medium towards globular clusters M 15 and M 92 \citep{Meyer99,Andrews01} and open clusters $h$ and $\chi$ Per by \citet{Points04}.
Complex and significant variations are often found down to spatial scales of 70,000 AU. 

While spatial variability of optical profiles is common,
the number of stars that have shown temporal variability is relatively small.
For example, 
\citet{Lauroesch02} found that only $\sim15$\% of stars in their sample showed temporal variability. In several directions with observed temporal variability densities of 20--200 cm$^{-3}$ were estimated, again signaling complex underlying physical conditions instead of changes in the total column density that could be caused by a passage of a discrete interstellar cloud. In addition, most temporal variations were observed in trace species and the majority of sightlines were behind supernova remnants (SNRs) or  HI shells. 


\subsubsection{Studies with larger samples}
More recent studies of TSAS have focused on expanding sample size, as well
as observing multiple ionization species so that electron density and the total hydrogen number density can be 
estimated. Following \citet{Danks01} who studied 10 stars 
and \citet{Smoker11} who studied 46 sightlines (with only 3 showing temporal variability),
\citet{McEvoy15} represents the largest sample of stars searched for absorption
 variability over two epochs. 104 stars in total were studied over periods 5-20 years (covering a range of spatial scales of a few to $\sim200$ AU). However, only 6\% of stars 
 showed variability in Na I, Ca I or Ca II.
 The study concluded that while less common at $<360$ AU, TSAS could be still ubiquitous on scales $>480$ AU. By assuming Ca ionization equilibrium, a temperature of 100 K, 
and that Ca I and Ca II sample the same line of sight, they estimated 
$n_e$, and then $n_H$.
All sightlines with detected variability (6\% of the total sample) have been found to have $n_H>1000$ cm$^{-3}$, in agreement with typical TSAS structures.
While the Ca ionization equilibrium is a promising technique for estimating $n_e$,
and confirming TSAS based on $n_H$, this study found $n_e=0.2-0.6$ cm$^{-3}$
which is higher than typical estimates from other dominant species.
In addition, several previous studies have shown that Ca estimates of $n_e$ appear to be 
significantly overestimated. This is a concern and likely suggests that 
assumptions used in the Ca ionization equilibrium  calculations
need to be revisited. As we summarize in the next section, many cases of optical variability have been found close to SNRs or expanding shells. As these are highly dynamic regions, often dominated by shock chemistry, the assumptions of photoionization equilibrium may not be appropriate.

\begin{figure}
\includegraphics[scale=0.7]{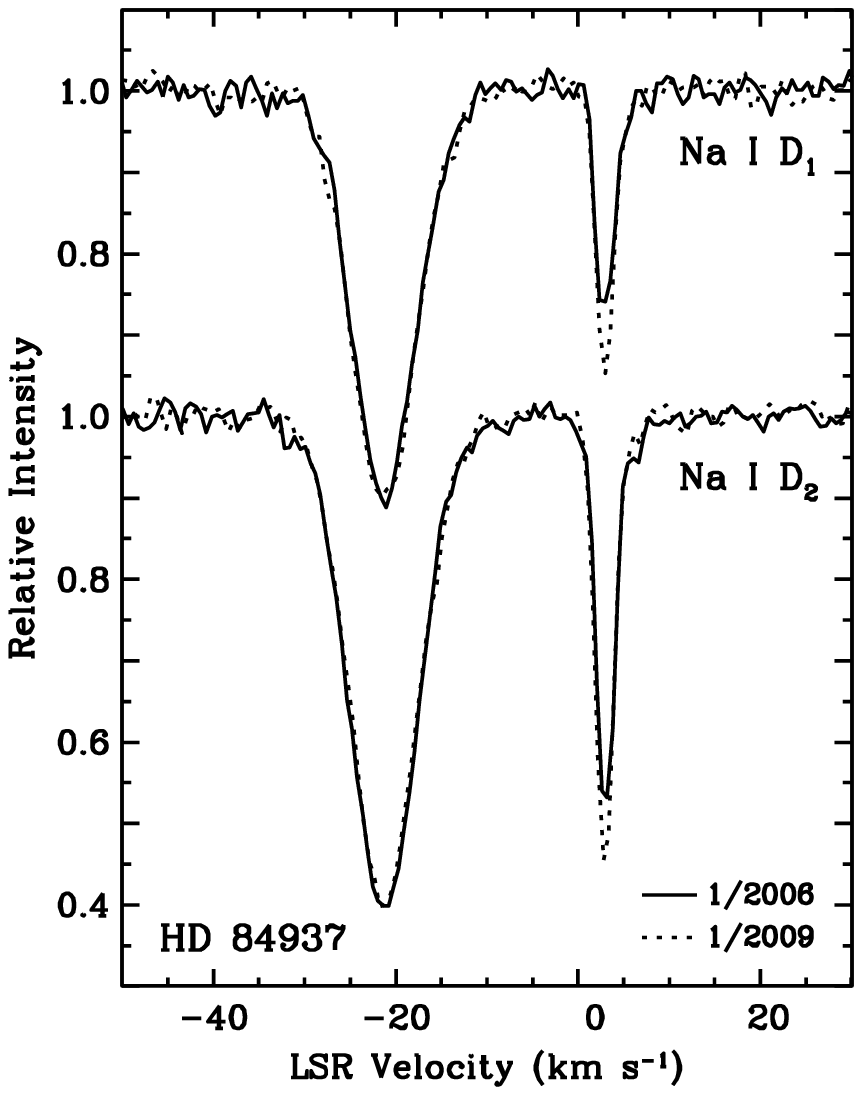}
\caption{Comparison of the NaI
D absorption profiles toward HD 84937 in KPNO coude feed spectra taken in 2006 January and in 2009 January. 
While the broader stellar NaI 
absorption at $−21$ km s$^{-1}$
is identical in both spectra, the narrow LLCC NaI
absorption exhibits a significant difference in
strength between the two epochs. From Meyer et al. 2012. }
\label{fig1}
\end{figure}


An even larger sample of 800 sightlines was obtained by \citet{vanLoon13}.
These authors used the Very Large Telescope to observe stars in the direction of 30 Doradus in 
the Large Magellanic Cloud, focusing on  4428 and 6614 \AA~ DIB features, as well as
Na I and covering  both LMC and Galactic velocities.
While the origin of DIBs is still not known, it is accepted that they 
trace diffuse ISM \citep{Herbig93,vanLoon13}.
By looking at the difference between pairs of Na I spectra, van Loon et al. (2013) noticed significant 
variations that gradually increase over the spatial scales probed, 2000 to 35,000 AU, 
assuming that Galactic absorbing gas is at a distance of 100 pc.

In summary, most recent optical studies, with larger samples and direct 
estimates of the TSAS density,  suggest that only a few percent of sightlines 
exhibit TSAS-tracing variability.

\subsubsection{TSAS observations in the direction of  supernova remnants}

Since early days optical TSAS examples were occasionally found in the direction of
SNRs and many studies have specifically searched  
for optical variability towards sources within and
in the vicinity of supernova remnants. In particular, Vela has been extensively studied 
\citep{Hobbs91,Danks95,Cha00,Welty08}
and several exceptional cases of optical variability have been discovered.

More recently, \citet{Rao16} studied Na DI spectra of 64 OB stars 
in the direction of Vela and made a temporal comparison
with earlier observations by  \citet{Cha00}. 
For three of the stars that showed major decrease 
in the low-velocity absorption, \citet{Rao17} obtained also SALT Ca II K absorption spectra
and found lack of variability.
While it is still puzzling how can Na I and Ca II profiles be so different, 
this study suggests that shocks associated with the SNR could be 
destroying local clouds and causing optical variability. Similarly,
\citet{Dirks16} studied HD47240, a star located behind 
the Monoceros Loop SNR over a period of 8 years, 
probing angular separations of $\sim10$ AUs. They found dramatic changes in Na I profiles and 
even a splitting of one velocity component, and  concluded that such drastic variability is 
related to local phenomena.
and likely not a pervasive component of the ISM. 

Time variability toward $\kappa$ Velorum was studied in \citet{Smith13} and a 
continuous increase in the equivalent width  and column density of Ca I and K I lines was found 
on scales 5-25 AU, but not in several other species. With estimated depletion pattern and electron density they calculated  $n_H>7 \times 10^3$ and $2 \times 10^4$ cm$^{-3}$. 
Based on distance constraints, it is likely that the absorbing clouds are located just beyond the edge of the Local Bubble. 

\citet{Meyer12} studied the Local Leo Cold Cloud
(LLCC) located inside the Local Bubble and at a distance 11--24 pc.
Using  C I lines, they  measured very high thermal pressure of $P/k\sim 6\times 10^4$ K cm$^{-3}$.
With an estimated temperature of  20 K and the HI column density of $10^{19}$ cm$^{-2}$, these observations implied $n_H\sim3000$ cm$^{-3}$ and a line-of-sight thickness of 200 AU. 
Multi-epoch Na I observations in the direction
of several stars found evidence for 46-AU structure in the direction of one star,
while no change in several stars which have much smaller proper motion.
They suggested that such small, over-dense TSAS clouds  inside the Local Bubble could be formed 
at  a collision interface between flows of warm clouds. 


In summary, many optical studies have 
suggested a possible connection between SNRs and TSAS traced by optical observations, e.g. \cite{Crawford03}.
\citet{Meyer15} found Na I variations in 12/20 sightlines with all cases being associated with SNRs or stellar bow shocks. 
These results suggest that physical processes associated with shock propagations and SNR evolution are important for TSAS formation and survival. 
This conclusion agrees with Stanimirovic et al. (2010) who found persistent HI optical depth
variability in the case of PSR B1929+10, which is likely tracing TSAS formed in fragmentation of the Local Bubble wall.
Motivated by the theory of evaporation
of cold clouds in a {\textit{hot}} medium \citep{Cowie75,Cowie77},
Stanimirovic et al. (2010) suggested that while similar fragmentation events occur 
frequently throughout the ISM,
the warm medium surrounding these cold cloudlets  induces  a natural selection effect
wherein small TSAS clouds evaporate quickly and are rare,
while large clouds survive longer and become a general property
of the ISM. This is discussed further in \S\ref{s:neutral_theory}.

\subsection{Theory of Neutral Structures}
\label{s:neutral_theory}

The main issues in the theory of TSAS are how structure forms on such small scales, whether the structures are as highly overpressured as implied by simple analysis,
and whether they are in equilibrium or highly transient. Addressing these issues requires accounting for a wide range of physical processes 
including turbulent flows, magnetic fields, radiative and conductive cooling,  ion-neutral friction, and possibly self gravity.
In addition, for both electrons and atomic tracers there is  continuing debate 
over whether TSAS and TSIS exist throughout the 
ISM or whether they are produced primarily in ``special'' locations such as shocks and shells.  
Whether the sites of TSAS and TSIS are physically associated is 
an open question that could be addressed by better statistics and coordinated measurements.

In this section, we first consider the possibility that TSAS could be related to
interstellar turbulence, and then consider implications for TSAS if it is in the form of tiny clouds.



\subsubsection{Interpretation of optical depth statistics}
\label{s:turbulent-origin}

An alternative to the picture of TSAS as tiny overpressured cloudlets was presented in
\cite{Deshpande00b} and is based on the analysis of HI optical depth statistics discussed in \S 2.1.4.

Spatial variation of optical depth $\tau(\mbfx)$ can be quantified by the structure function: 
%
\begin{equation}\label{eq:tau}
S(\Delta\mbfx)=\langle\vert\tau(\mbfx)-\tau(\mbfx + \Delta\mbfx)\vert^2\rangle.
\end{equation}
If $S(\Delta\mbfx)$ is a power law, $S(\Delta\mbfx)\propto\vert\Delta x\vert^{\alpha}$, then, from \citet{Lee75}, 
%
%
the optical depth variation $\Delta\tau(\Delta x)$ should scale as $(\Delta x)^{(\alpha - 2)/2}$.
For $\alpha = 2.75$, the value measured for Cas A \citep{Deshpande00b}, $\Delta\tau(\Delta x) \propto\Delta x^{0.38}$. This means that  
blotchy optical depth at large scales translates to significant opacity variations at small scales. An oversimplifed but suggestive way to visualize this is to imagine looking through a set of large,
semitransparent disks suspended randomly in space against an illuminated background. There will be some closely spaced pairs sightlines that pass through different numbers of disks and therefore
see different levels of brightness. However, this small scale variation is produced by structures much larger than the separations of the sightlines. 

Thus, \cite{Deshpande00a} argues that the observed small scale variations in $\tau$ do 
not require tiny, overdense, overpressured clouds. Rather, 
 the observed optical 
depth differences are consistent with
a single power-law description of the HI optical depth distribution as a 
function of spatial scale. 
The observed variations in optical depth sample
the square-root of the structure function of the HI optical depth, not directly the power 
spectrum of optical depth fluctuations. This is a key point.
As to
 why the autocorrelation of the HI optical depth should have the power law property, and what controls $\alpha$ we turn to theories of turbulence.
 
\subsubsection{Turbulence in the neutral ISM}

Turbulence in neutral gas can produce small scale structure in two different ways. 
Interstellar turbulence, being transonic or mildly supersonic, 
is compressible. Density fluctuations are a natural part of 
compressible turbulence, and, if the turbulence is subsonic or transonic, have a spectrum similar to the energy spectrum (\S 1.3).The only 
question is whether compressible turbulence can persist to AU scales, or becomes noncompressive 
or  dissipates on larger scales. Small scale density fluctuations are also generated from large scale density gradients by chaotic  fluid motions, which bring
previously widely separated fluid elements, with very different densities, into close
proximity.

Turbulence in the CNM is both magnetized and weakly ionized. Magnetized turbulence in weakly ionized gas is discussed in 
\cite{Cho02,Cho03,Li06,Inoue08,Tilley11,Burkhart15,Zweibel15}. The 
magnetic field acts directly only on the plasma, which transmits magnetic forces to the neutrals through collisions on a timescale $\tau_{ni}\equiv\nu_{ni}^{-1}\sim 5.3\times 10^8/n_i$ s \citep{Draine83} for an ion number density $n_i$ and ions more massive than neutrals. If $\tau_{ni}$ is short compared with the characteristic dynamical time $\tau_{dyn}$, the plasma and  
the neutrals move together as a single, magnetized fluid, while for $\tau_{ni}/\tau_{dyn}< 1$,
they decouple. For processes on scale $L$ we define $\tau_{dyn}\equiv (L/v_A)(1+M_A)^{-1}$,
where $v_A\equiv B/\sqrt{4\pi\rho}=2.0\times 10^5B_{\mu}/n_n^{1/2}$ is the Alfven speed in cm s$^{-1}$, $B_{\mu}$ is magnetic field strength in $\mu$G, $n_n$ is the neutral number density, and $M_A\equiv v(L)/v_A$ is the Alfven Mach number defined with respect to the typical velocity at scale $L$. We then define the
decoupling scale $L_d$ as the scale at which $\tau_{dyn}=\tau_{ni}$:
\begin{equation}\label{eq:Ld}
L_d\equiv\frac{v_A\tau_{ni}}{1+M_A}  \sim 1.1\times 10^{14}\frac{B_{\mu}}{n_n^{1/2}n_i(M_A+1)}{\rm{cm}}.
\end{equation}
$L_d$ is the scale below 
which the magnetic field and plasma slip through 
the neutral fluid in less than the flow time. Equation~\ref{eq:Ld} agrees with the criteria given by
\citet{Klessen00}, \citet{Zweibel02}. This is also the scale at which  the ambipolar Reynolds number $R_{AD}$ defined in \citet{Zweibel97} is of order unity. In the cold, neutral ISM, 
$L_d$ is typically a few tens to hundreds of AU. 

At scales below $L_d$ the plasma and neutrals follow different dynamics. The plasma develops 
an independent cascade with many current sheets  and associated electron density filaments that
if sufficiently dense could be an important source of interstellar scintillation.  The neutrals 
evolve independently of the plasma and magnetic field. Since $L_d$ is generally much less than the outer
scale of turbulence, the hydrodynamic cascade shortward of $L_d$ is expected to be quite subsonic 
and therefore incompressible. Density fluctuations become large only if the medium is nearly
thermally unstable, in which case large, early isobaric density fluctuations could occur \citep{HennebelleAudit07}. However, a calculation 
with a large dynamic range and all physical 
processes accounted for is still lacking. 
Notably, however, the slope of the power spectrum of 
the optical depth autocorrelation function is close to 
the prediction of \citet{Lazarian04} for optically thick media. 

Large pressure fluctuations are possible in a turbulent medium; they tend to be associated with large velocity fluctuations. 
As a rough guess, we estimate that to confine a structure
that is overpressured by a factor of 100, a turbulent velocity at least 10 times the rms velocity is required \citep{McKee92}.  If the velocity probability density function (PDF) is Gaussian, the probability of creating such a velocity is exponentially small,
but if the turbulent velocity PDF has  a power law tail the fraction of power at high velocities is larger. For example, \citet{Falkovich97} suggested that the velocity PDF has the form $(v/v_0)^{-4}$ for $v/v_0\gg 1$. Non-Gaussian velocity PDFs could, for example, result from stellar winds, self-gravity, or even large-scale turbulent driving \citep{Ossenkopf02}.
While density PDFs in turbulent flows have been studied by many authors, and velocity PDFs have been studied for self gravitating turbulence \citep{Klessen00}, characterizing the velocity and pressure PDFs at small scales, including cooling, is an important problem for the future. On the observational side, better statistics on the line of sight and covering factor distributions would tighten the constraints on turbulence models of pressure fluctuations. 

\subsubsection{Tiny neutral clouds}\label{s:tiny} 

Having considered the possibility that observations of TSAS could be explained by a power spectrum of opacity variations, and discussed their relationship to
interstellar turbulence, we now discuss whether they could, in fact, be tiny clouds, as proposed originally.
If the clouds are spherical, they must be highly overpressured. In \S\ref{s:pressure}, we estimated the energy input to the ISM resulting from free expansion of tiny overpressured clouds, and argued that this sets an upper limit on the filling factor of such clouds. The overpressure problem is less extreme for elongated clouds viewed end on, and we account for this
in discussing gravitational and magnetic confinement.

{\bf Thermal equilibrium.}
We assume the clouds are heated by photoelectrons from dust, as in the rest of the CNM. Then, from \cite{Wolfire03}, especially Figures 8 and 11, we see that it is possible to maintain thermal equilibrium for high densities and
low temperatures, e.g. $n_H \sim 10^3$ cm$^{-3}$, $T\sim 30$ K, similar to what was derived for the Local Leo Cold Cloud \citep{Meyer12} and warmer than proposed in \cite{Heiles97}, which invoked cooling by molecular species as well as by the C II 158$\mu$m line, which normally dominates CNM cooling. The resulting
pressures are still high, and so in the absence of a confinement mechanism the clouds will expand. 

Assuming that the C II fine structure line is the
main source of radiative cooling, the radiative loss function is \citep{Spitzer78} $\Lambda = 7.9\times 10^{-27}d_Ce^{-92/T}$ cm$^3$ s$^{-1}$, where $d_C$ is the carbon depletion factor. 
The result agrees with the fitting formula in \cite{Wolfire03} to 20\% at 100 K if we take $d_C=0.35$
to match the gas phase $C$ abundance they assumed,
and can be extrapolated to lower temperatures. Estimating the dynamical time $t_{dyn}$ as $R/c_s$, where $c_s\equiv\sqrt{kT/m}$ and the cooling time $t_{cool}\equiv 3kT/2n\Lambda$ we define the minimum column density $N_{min,c}$ such that $t_{dyn}/t_{cool} > 1$:
\begin{equation}\label{eq:numfield}
N_{min,c}=1.2\times 10^{15}T^{3/2}e^{92/T}{\rm{cm^{-2}}}.
\end{equation}
For $T=50$ K, eqn, (\ref{eq:numfield}) gives $2.7\times 10^{18}$ cm$^{-2}$, which is in the range for TSAS (see Tables 1 and 2). We might expect clouds
with $N\ll N_{min,c}$ to expand (and cool) adiabatically. Clouds with $N\gg N_{min,c}$ would cool radiatively faster than they expand and might then reach a thermal equilibrium - which is still overpressured.

{\bf Gravitational confinement.}
Overpressured clouds can maintain equilibrium if they are gravitationally or magnetically
confined. 
For a given column density, self gravity is
less effective relative to external pressure in confining a prolate cloud than a spherical cloud.
It can be shown from the virial theorem, using expressions in \cite{Bertoldi92}, that
the ratio of the external pressure confinement term to
the gravitational confinement term $X_c$ is:
\begin{equation}\label{eq:Xc}
X_c=\frac{15P_s}{\pi GN^2m_H^2}\frac{y\sqrt{y^2-1}}{\ln{\left(y+\sqrt{y^2-1}\right)}},
\end{equation}
where $N$ is the column density along the major axis, $P_s$ is the surface pressure, and $y$ is the ratio of major to minor axis. 
Equation (\ref{eq:Xc}) shows that gravitational confinement of highly elongated clouds ($y\gg 1$) is much less effective relative to pressure confinement than it is for spherical clouds ($y = 1$). For example,
if $y=10$, $X_c$ is about 33 times larger than for  a spherical cloud. Therefore, filamentary clouds must be very close to pressure balance with their environments, even accounting for self gravity.


{\bf Magnetic confinement.}
Overpressured filamentary clouds can maintain equilibrium if they are 
magnetically confined. Many
configurations are possible \citep{Fiege00}, but here we consider the simple case of a filamentary cloud of radius $R$ with internal pressure $P(r)$ confined by a helical magnetic field $\mbfB=B_z\hat z + B_{\phi}(r)\hat\phi$, with $B_z$ a constant. The magnetic force points radially inward if $rB_{\phi}$ is an increasing function of $r$. It can be shown from the virial theorem that the mean internal cloud pressure $\langle P\rangle$, external pressure at the cloud surface $P_s$, and azimuthal magnetic field component at the cloud edge $B_{\phi}(R)$ are related
by:
\begin{equation}\label{eq:magneticconfinement}
\langle P\rangle=P_s+\frac{B_{\phi}^2(R)}{8\pi}.
\end{equation}
(The axial component $\hat z B_z$ drops out because it is the same inside and outside the cloud). If $P_s$ is of order the ambient ISM pressure, then the pressure in $B_{\phi}(R)$ must
be comparable to the TSAS pressure. For example, assume $B_{\phi}=B_0r/R$ for $r<R$ and $B_{\phi}=B_0R/r$ for $r>R$. In order to satisfy force balance, $P=P_c-B_0^2r^2/(4\pi R^2)$ for
$r < R$ and $P=P_c-B_0^2/4\pi$ for $r > R$. The ratio of surface pressure $P_s=P_c-B_0^2/4\pi$
to central pressure $P_c$ is $1-B_0^2/4\pi P_c$, which is small only when $B_0^2$ is slightly less than $4\pi P_c$.
A magnetic field of 10$\mu$G (about twice the rms galactic field; \cite{ZweibelHeiles97}) would
support a central pressure $P_c/k=5.8\times 10^4$ K cm$^{-3}$.

A cloud can only be magnetically confined for as long as it takes the magnetic field and plasma component to drift through the neutral component, which comprises most of the mass. In the situation described here, the
magnetic field and plasma drift inward, which allows the neutral component to drift outward.
From \S 2.4.2 (see eqn. (\ref{eq:Ld})), and assuming rough equality of gas and magnetic pressure,
we can write the ratio of the drift speed $v_D$ to the expansion speed $c_S$ as:
\begin{equation}\label{eq:vdcs}
\frac{v_D}{c_S}\sim\frac{c_S\tau_{ni}}{R}\sim\frac{\tau_{ni}}{\tau_{dyn}}\sim 3.5\frac{v_5}{n_iR_{AU}},
\end{equation}
where $v_5$ and $R_{AU}$ are the sound speed and radius in km/s and AU, respectively. Equation
(\ref{eq:vdcs}) says that the cloud can be confined for more than its free expansion time
if the neutral-ion collision time is short compared to the dynamical time, and sets a lower
limit on the radius of a magnetically confined filament. For example, if $n_i = 0.1$ cm$^{-3}$
(ionization fraction of 10$^{-4}$ with $n=10^3$ cm$^{-3}$) and $v_5=1$ km/s, then $v_D/c_S < 1$ for $R > 35$
AU.
%
%
The similarity of this scale to TSAS
suggests that there is a critical width below which filaments cannot be magnetically confined,
%
%
whereas for larger TSAS, magnetic confinement is possible for quite long times. The filamentary morphology is potentially attractive for explaining thin H I fibers \citep{Clark14,Kalberla16}.

{\bf Thermal conduction.}
As noted by \cite{Heiles97}, the overpressure problem is reduced if TSAS is colder than the surrounding CNM. Heat will then flow into the cloud, which will either shrink due to evaporation or grow due to condensation, depending on whether the column density is
below (evaporation) or above (condensation) a critical column density 
 $N_{c}$  at which radiation and conduction balance \citep{McKee77}. Balancing the conductive 
heating and radiative cooling terms in the energy equation yields \citep{Inoue06}:
\begin{equation}\label{eq:field}
N_{c}=\left(\frac{\kappa T}{\Lambda}\right)^{1/2},
\end{equation}
where $\kappa$ and $\Lambda$ are the thermal conductivity and radiative loss function, respectively. For the CNM we take $\kappa = 2.5\times 10^5T^{1/2}$ erg s$^{-1}$ K$^{-1}$ cm$^{-1}$
and $\Lambda = 2.8\times 10^{-27}e^{-92/T}$ cm$^3$ s$^{-1}$, corresponding to C II cooling with a gas phase Carbon abundance of $1.4\times 10^{-4}$. Substituting these expressions into eqn. (\ref{eq:field}) and normalizing $T$ to 30 K gives
%
$N_{c}=5.6\times 10^{17}T_{30}^{3/4}e^{1.53/T_{30}}{\rm{cm^{-2}}}$,
%
which is compatible with column density measurements for TSAS. 

Clouds with $N\ll N_c$ will evaporate on a timescale $\tau_{evap}$:
\begin{equation}\label{tauevap}
\tau_{evap}\sim\frac{3nk_BR^2}{2 \kappa}\sim 4.5 \times 10^{-9}\frac{NR_{AU}}{T_{30}^{1/2}}.
\end{equation}
 Clearly, conduction sets a lower limit on the thickness of a cold filament, which makes it difficult to maintain TSAS temperatures that are lower than other CNM temperatures and thus may explain why Brogan et al. (2005) found no
difference between TSAS and other CNM temperatures. On the other hand, larger structures will grow as they accrete condensing material.

\subsection{Summary and Outstanding Questions}

{\bf Ubiquity.} Spatial and temporal variability of different tracers, based on several different observational techniques, have found clear TSAS examples. However, as Tables 1 and 2 show, since early days TSAS was not seen in all observed directions. The preponderance of evidence is that TSAS is not ubiquitous in the ISM. 

It is important to keep in mind that the three categories of observational techniques we discussed in this section, each have their own pros and cons, and often probe very different spatial scales (e.g. Figure 1). The interferometric imaging of $\Delta \tau$ against extended background sources has a major advantage of providing spatial (and kinematic) 2D information about TSAS, and may even be able to constrain TSAS morphology in the future. Especially powerful are statistical measurements of the power spectrum of $\Delta \tau$. However, such measurements require exceptional sensitivity and are limited by the available source size, source surface brightness, and telescope resolution. Temporal variability of absorption profiles is more easily accessible observationally, yet suffers from one-dimensional sampling of the $\Delta \tau$ distribution. As summarized in this section, some tension exists between results from two categories of radio measurements; while interferometric imaging of 3C138 and 3C147 claimed pervasive TSAS, temporal radio variability against pulsars and point-source extragalactic targets suggests that TSAS is not a pervasive component of the ISM. The later conclusion is in agreement with optical studies of TSAS where abundant evidence exists of TSAS associated with SNRs.
Future high-cadence measurements are needed to resolve this tension.
Brogan et al. (2005) questioned the ability to constrain the TSAS filling factor via 
temporal variability of absorption profiles as often a small number of observing epochs is used and the observational cadence is rarely matched to the size scales of $\Delta \tau$ variations. On the other hand, optical temporal or spatial variability measurements have a strong tracer dependence, and require multiple transitions for $n_e$ estimates, and often many complex equilibrium assumptions.


{\bf Volume filling factor.}
The TSAS detection rate is directly related to the estimates of the volume filling factor. As we showed in this section, the heating rate via structure expansion that results from over-pressured TSAS has significant consequences on many areas of astrophysics, including dust and gas temperature, and places an upper limit on the overpressured TSAS volume filling factor.
For example, the volume filling factor of TSAS of $\sim1$\%, which is 
close to estimates from interferometric imaging of 3C138 and 3C147, would imply a heating rate that is at least 1000 times larger than the heating due to photoelectric effect. On the other hand, McEvoy et al. (2015)'s detection rate of 6/104 sightlines, translates to a volume filling factor of $\sim 10^{-6}$ (if we assume that a typical line of sight is 100 pc and TSAS is about 100 AU in size). This is more reasonable regarding the implied heating rate. Future constrains of the filling factor are very important, as filling factors close to 1\% would imply potentially the existence of unknown cooling processes in the ISM and/or that TSAS is either not overpressured or somehow confined.

{\bf Likely sites of TSAS and formation mechanisms.}
Especially over the last decade,
abundant observational evidence has emerged (using optical and radio observational techniques) that TSAS is frequently found close to SNRs. This suggests that shocks and SNR evolution (via shell fragmentation and/or cloud destruction processes)
is of high importance for TSAS formation, survival, and integration with the surrounding ISM. 
Besides turbulent ISM simulations that rarely reach down to AU scales, only two numerical studies have addressed TSAS formation.
\cite{HennebelleAudit07} simulated collision of WNM flows and the formation of CNM clouds. The collision of turbulent WNM forms a thermally unstable region with enhanced cooling, which fragments, forming small clouds. They found that the CNM fragments collide with each other, forming transient, shocked regions with properties similar to those of TSAS. In their 2D simulation, only 0.1\% of lines of sight cross fragments with $n>10^4$ cm$^{-3}$. \cite{Koyama02} performed a 2D hydrodynamical simulation of the WNM compression by supernova explosion and subsequent formation of small clouds. Similarly, the post-shock layer becomes thermally unstable and cools fast, forming CNM clouds on thousands AU scales, which eventually get embedded in the hot high-pressure gas.


\begin{figure}
\includegraphics[scale=0.5]{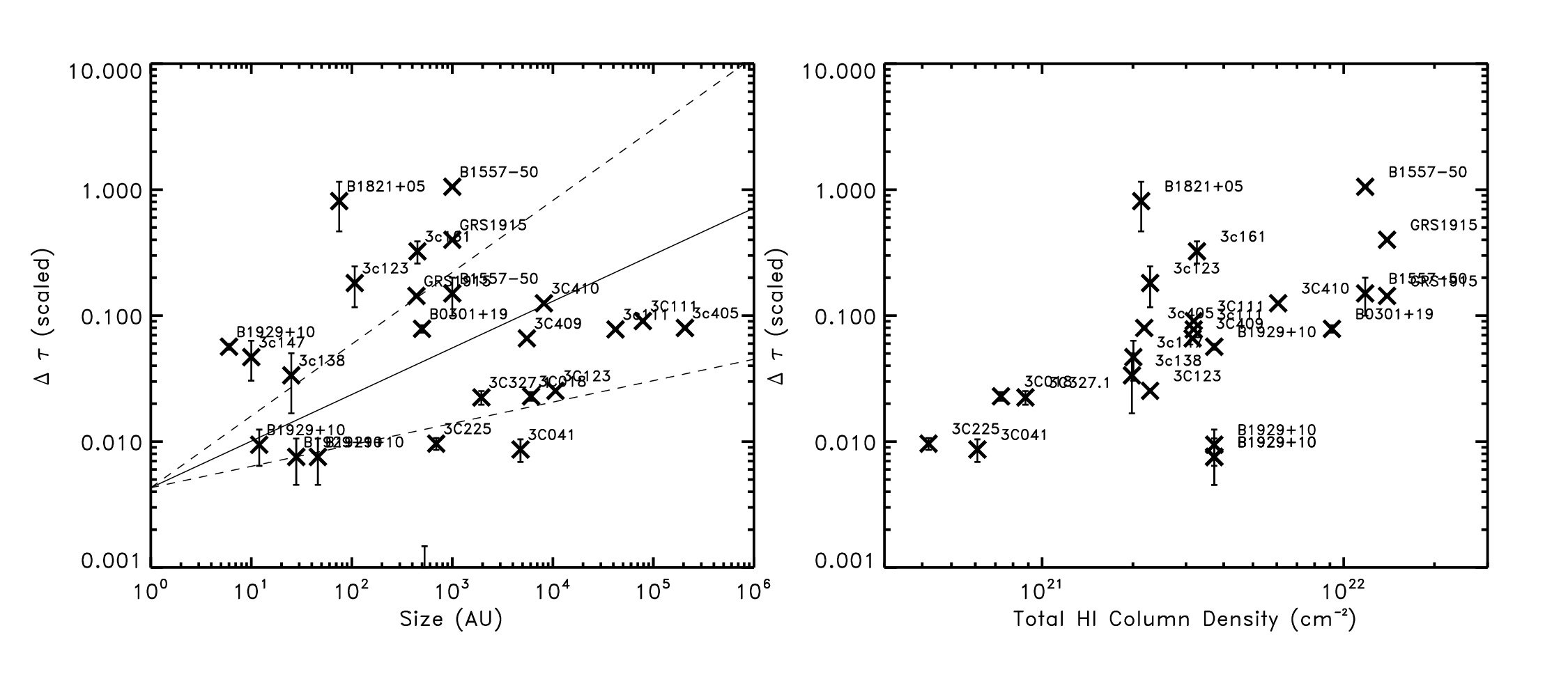}
\caption{(Left) Variations in HI optical depth as a function of spatial scale. Detections from Tables 1 and 2 are plotted after applying scaling to bring all $\Delta \tau$ values to the common velocity resolution (see text). The solid line shows the level of fluctuations in $\Delta \tau$ as predicted by Deshpande (2000b), dashed lines show $\pm2$-$\sigma$ slope uncertainties. (Right) Variations in HI optical depth as a function of the total HI column density along the line of sight. As explained in the text, the HI column density comes from several different surveys.}
\label{f:compare_desh}
\end{figure}

{\bf What is the minimum size of TSAS?}
As we showed in \S\ref{s:tiny}, the spatial scales of a few tens to hundreds of AU are very interesting as corresponding to the scales where turbulent dissipation is expected to happen, including viscous dissipation, which is expected to be the main dissipation mechanism for
hydrodynamic CNM turbulence \citep{Zweibel05}.
This is also the range of scales at which the magnetic field, plasma, and neutrals transition from strongly to weakly coupled. In essence, the astrophysics on TSAS-scales is rich and exotic, and
largely unexplored.

The turbulent power spectrum that exists in the ISM on larger scales is expected to continue below scales of hundreds of AU, however it becomes more complex as neutrals and ions split into different turbulent cascades.
On theoretical grounds, we expect that the neutral turbulent spectrum at these scales becomes incompressible and has a Kolmogorov power spectrum slope. However, 
thermal instability can modify the turbulent cascade and enhance density fluctuations
\citep{Hennebelle07}, while optical depth effects change the relationship between density and optical depth fluctuations, resulting in a  shallower power spectrum of optical depth. 
Theoretically, therefore, TSAS could be part of the turbulent spectrum but observationally this is still not confirmed and it is not clear that density and pressure fluctuations of the required amplitude can be achieved by typical ISM turbulence. Mapping of HI absorption against three extended sources (Cas A, Cygnus A and 3C138) showed a similar power spectrum slope of $2.7$ over a range $\sim10$ to 150 AU, and $\sim15,000$ AU to 3 pc. While this is impressive, we need to bridge the gap in the middle with future observations.
In addition, the power spectrum amplitudes for different regions are different.
The compilation of heterogeneous  observations shown in Figure~\ref{f:compare_desh} (left) do not show a continuous spectrum. Therefore, we do not have evidence for a {\it continuous} power spectrum from few pc all the way to few AU.

Several physical processes place a lower limit on the cold TSAS cloud size of $\sim100$ AU.  
Thermal conduction suggests that smaller TSAS clouds would quickly evaporate, while
magnetic field could not provide confinement for TSAS clouds smaller than $\sim100$ AU.
All of this suggests that, theoretically, TSAS on scales $<100$ AU should be short-lived and transient, while larger TSAS clouds should be more frequent.
Constraining observationally abundance of small vs larger TSAS clouds could constrain these important physical processes.

Probing spatial scales of $<100$ AU observationally requires high resolution interferometric observations, and/or high sensitivity and frequent temporal sampling of absorption spectra (radio and optical). Several experiments have detected TSAS on spatial scales of few to 100 AU, e.g. \cite{Brogan05,Stanimirovic10,Meyer12}, however
more detections exist on scales $>100$ AU. While scales $>100$ AU are observationally easier to probe, and a higher frequency of TSAS could be driven by observational biases,
the existing samples are very small and improving statistics is of high importance.

{\bf What defines the level of observed optical depth variability?}
While most sources showed $\Delta \tau <0.1$,
3C138 and 3C147 stand out as sources that have shown a much higher level of variability measured persistently for over three decades in scale. 
In addition, several optical studies have noticed an increase in absorption line variability on large spatial scales \citep{Meyer90,Meyer94,vanLoon13}. 
If $\Delta \tau$ is driven by interstellar turbulence, a power-law relation is expected
between TSAS size and $\Delta \tau$ as a consequence of the turbulence cascade from large to small scales. Based on the measured power-law of $\Delta \tau$ in the case of Cas A, Deshpande (2000a) predicted $\Delta \tau \propto l^{(\alpha-2)/2}$, with $\alpha=2.75$ and $\Delta \tau=0.2-0.4$ on scales 50-100 AU.
With larger statistical samples of TSAS we can start 
testing the predictions for both the slope and amplitude of the turbulent spectrum, as well as investigate possible
correlations between $\Delta \tau$ and other physical parameters (e.g. TSAS size, the total HI column density or the H$_2$ fraction). Clearly, a lot more on this will come in the future with larger samples.

In Figure~\ref{f:compare_desh} we show the detected level of
optical depth variations, $\Delta \tau$, from Tables~\ref{t:TSAS} and \ref{t:TSAS2}, as a function of TSAS spatial scale (left) and the total line-of-sight HI column density (right). We use here only radio detections and omit optical observations due to the complexity associated with non-dominant ions.
While $\Delta \tau$ is a key observable obtained directly from optical depth profiles, we note that spectral velocity resolution varies among observations (we have listed this information in Tables 1 and 2). To compare all measurements we scaled $\Delta \tau$ to the common velocity resolution of 7.3 km/s (this is the lowest velocity resolution in the sample). The total HI column density along lines of sight toward background sources was obtained either from individual references provided in the tables, or from the Arecibo Millennium (Heiles \& Troland 2003) and 
21-SPONGE HI absorption-line surveys (Murray et al. 2017, in prep) if this information was missing. These measurements include a correction for high optical depth. For a sub-sample of sources (3C147, 3C161, GRS1915, 3C405, PSR B1821+05, PSR B1557-50, PSR B0301+19) we did not have the full column
density information and have obtained integrated HI column densities from 
the Effelsburg all-sky, HI emission survey \citep{HI4PI16} by using the optically-thin limit. These are therefore lower limits of the HI column density. 

The left figure is essentially a scatter plot, there is no obvious correlation between the observed optical depth variation and spatial scales for measurements sampling diverse interstellar directions. 
As discussed in Section~\ref{s:neutral_theory}, 
interstellar turbulence would result in a correlation between
$\Delta \tau$ and linear scale and for Cas A Deshpande (2000a) measured
 $\Delta \tau \propto l^{0.38}$ (Roy et al. 2012 confirmed the same slope for 3C138 over the range of scales 5--100 AU). We overplot the predicted level of variations from Deshpande (2000a)
in Figure~\ref{f:compare_desh} as a sloping line, with dashed lines corresponding to $\pm2$-$\sigma$ slope uncertainties. As with observations, we have scaled the predicted $\Delta \tau$ to correspond to a velocity resolution of 7.3 km/s. The lack of correlation in this plot may suggest that thermal conduction and/or magnetic field-gas decoupling are more important than turbulence in shaping the ISM structure on AU-scales. However, improving statistics is essential to understand processes at AU scales, especially by focusing on individual regions with a single set of physical parameters such as the energy input. 


The right-hand figure appears more interesting as it suggests a rough correlation between $\Delta \tau$ and the total HI column density.
Galactic directions with higher column density appear to
have higher level of $\Delta \tau$ variations. 
This correlation follows the Galactic latitude trend:
at $|b|>25$deg, $N(HI)<10^{21}$ cm$^{-2}$ and $\Delta \tau \sim 0.02$;
at $5<|b|>10$, $N(HI)=10^{21-22}$ cm$^{-2}$ and $\Delta \tau \sim 0.01-1$;
at $|b|<5$deg, $N(HI)>10^{22}$ cm$^{-2}$ and $\Delta \tau > 0.1$.
TSAS fluctuations appear larger at low latitudes.
The increase of the HI column density at low latitudes is a consequence
of the plane-parallel disk geometry, once column densities are corrected by a $\sin(b)$ factor they stay roughly constant \citep{Dickey90}.
The increase of $\Delta \tau$ with the column density
is in agreement with Frail et al. (1994) and Stanimirovic et al. (2010), who 
noticed that directions with high optical depth (and therefore high CNM column density) had higher level of $\Delta \tau$ variations.
TSAS seems to prefer lower Galactic latitudes and higher HI column density, similarly to the CNM, H$_2$ and SNR distributions, again potentially signaling preferred
formation sites.
This correlation also suggests that TSAS detections are intimately
associated with the larger scale structure of the parental HI clouds,
and not a totally random phenomenon. 
In addition, as hinted by \cite{Stanimirovic10}, $\Delta \tau$ increases with $\tau$ suggesting that the amplitude of density fluctuations is not universal as would be the case for universal interstellar turbulence. Instead, density fluctuations appear stronger at low Galactic latitudes and high HI column densities (and optical depths), consistent with the idea that excess energy injection occurs in such environments likely via supernova explosions.



\subsubsection{Outstanding Questions}

Several key questions concerning TSAS origin and properties remain open.

{\bf Turbulent power spectrum and dissipation scales.}
On the theoretical side, numerical simulations going all the way to dissipation scales and predicting the behavior of the power spectrum at such scales do not exist, and creating them would be a formidably computationally intensive task given the large range of space and time scales involved. Besides an empirically-driven power spectrum (based on Cas A, Cygnus A and 3C138 observations), we have no numerical/theoretical predictions for the level  of $\Delta \tau$ or column density variations that can be compared with observations. A related questions concerns velocity PDFs.
While density PDFs have been studied extensively numerically, very few studies of velocity PDFs exist. Characterizing velocity fluctuations, in theory and observations, is important as this can potentially drive pressure fluctuations. The anisotropy of fluctuations formed in this manner should be
carefully investigated to assess whether anisotropic turbulence can create the H I fibers mentioned
in \S 1.2 \citep{Clark14,Kalberla16}.

The spectacularly similar power-spectrum slope for Cas A, Cygnus A and 3C138 covers spatial scales of $\sim10$-150 AU and $\sim15,000$ AU to 3 pc. This is almost six orders of magnitude but with a gap in the middle. These results are very exciting as they could be shaping ``the Big Power Law in the Sky'' but for the cold, absorbing medium, which could have large implications for TSAS and heating/cooling in the ISM. 
Filling the gap on scales 150--15,000 AUs is essential, as well as understanding the variations of the power spectrum amplitude across different regions.
More interferometric mapping of extended radio sources is needed, using both SNRs and quasars. In addition, the two best studied extragalactic sources 3c138 and 3c147 both found TSAS only on spatial scales close to the resolution limit.  
Future observations are essential to confirm that this is not an observational artifact. Telescopes like MeerKAT will be able to test this against other SNRs.

As mentioned already, better statistics are essential to start making progress in understanding the origin and properties of TSAS. Existing observations span many observational techniques, each with its own systematics and limitations. Combining such diverse and small samples to search for global properties and trends is challenging. The related issue is inhomogeneous methodology used by different authors in estimating distance to the absorbing TSAS or spin temperature. 
Homogeneous samples could be achieved though frequent, long-term monitoring of sources (pulsars or extragalactic) with already detected TSAS. This would provide better sampling of individual lines of sight, enabling monitoring of individual TSAS features to see how they evolve with time. For example, pulsars with detected variability should be monitored to provide longer time baselines and test TSAS evolution (do certain directions show persistent TSAS, or does TSAS eventually disappear?).

{\bf Formation mechanisms.}
As an emerging trend points at supernovae as a potential common birth place of TSAS, detailed numerical simulations are needed with predictions for the TSAS spectrum and expectations for the frequency of TSAS away from birth places. It would also be worthwhile to isolate various critical processes in the turbulent cascade, such as the
evolution of entropy modes when cooling is included, ion-neutral decoupling, formation of vortex filaments that could wind up a magnetic field, and sites of enhanced magnetic dissipation
\citep{Zhdankin13, Zhdankin15, Zhdankin17}.
In addition, observational evidence that TSAS is less common away from SNRs is still lacking. Large-scale surveys could conclusively quantify TSAS abundance between localized regions and random ISM fields.

{\bf TSAS environment and internal structure.}
Long-term monitoring of directions with TSAS will enable studies of internal structure of TSAS. Observational constraints of TSAS spin temperature and magnetic field essentially do not exist and are key details for understanding TSAS over-pressure. For example, thermal conduction requires TSAS temperature to be similar to the CNM temperature. Similarly, theory suggests that larger TSAS can be easily magnetically confined with a magnetic field or order of 10 $\mu$m. These are clear theoretical predictions that need to be confronted observationally.  In addition, 
thermal equilibrium considerations suggest that TSAS clouds larger than $\sim100$ AU would cool radiatively. This would result in significant CII emission associated with TSAS that can be tested observationally.

At high TSAS densities it is possible to have various molecular species. This can be easily tested with the Atacama Large Millimeter Array (ALMA) observations of CO and other transitions by focusing on directions known TSAS. ALMA is also perfect for monitoring strong extragalactic sources and measuring molecular absorption from species like HCO$^+$, HCN, HNC etc which have been found in diffuse interstellar regions
in the Milky Way \citep{Liszt96}. Similarly, \cite{HennebelleAudit07} and \cite{Koyama02} suggest that TSAS could be formed in shocked CNM. Observing shock tracers like SiO with ALMA can test this possibility.


In addition, broader observed velocity range can test more extreme origin theories  ``halo''. 
Most observations focus on high velocity resolution and
do not have broad velocity coverage. 
Finally, the under- and over-pressure directions based on C I measurements by Jenkins \& Tripp (2011) 
are fascinating and require further investigation.

\section{Tiny Scale Ionized Structure}
\label{s:TSIS-general} 


The warm ionized medium contains density fluctuations over a broad range of spatial scales which provide a unique imprint on propagating radio waves.
Turbulent fluctuations in the plasma density and magnetic field cause stochastic spatial and temporal fluctuations in the refractive index in the ISM. As a result, radio waves propagating through a plasma medium experience variations.
Due to scattering of radio waves the flux density of pulsars and other pointlike sources varies with time and frequency
(interstellar scintillation), and angular size of radio sources increases due to the blurring effect of the turbulent medium. 
Through these effects,
observations of pulsars and pointlike
extragalactic sources have revealed
a spectrum of 
plasma structures in the electron density 
distribution
which follows a Kolmogorov spectrum over 5 or even 10 decades 
(Armstrong et al. 1995,
Haverkorn \& Spangler 2013). The inner scale of this turbulent cascade is 70-100 km \citep{Rickett95}, while the outer scale reaches 4-30 pc. Diffractive scintillation effects produced by small scale structure decrease with increasing frequency while the opposite is true for refractive scintillation produced
by larger scale structure \citep{Rickett84}.
The
outer scale seems to suggest 
the variance of density fluctuations
$\langle\delta n_e^2\rangle^{1/2}\sim 10^{-3}$.
At a WIM temperature of 8000 
K and electron density of $\sim0.3$ cm$^{-3}$,
these tiny-scale
density fluctuations do not pose the enormous overpressure problems
of TSAS or the phenomena discussed below. 
While the Kolmogorov spectrum persists over several orders of magnitude,
the line of sight distribution of 
the
turbulent material is still not well understood (uniform filling factor or confinement to thin layers?).
We return to this topic in \S~\ref{s:ionized_theory}.

However,
there are several observational phenomena that exhibit flux modulations and 
scattering angles that are much larger than what is expected for Kolmogorov turbulence. 
These involve: 
extreme scattering events, intra-day variability of flat-spectrum radio quasars, 
and parabolic arcs and arclets in pulsar ``secondary'' spectra. 
All these events are often referred to as ``extreme scattering'' and, assuming isotropy in shape (so the angular size on the plane of the sky can be combined with the column density to give volume density),  require plasma pressures 
that are much greater than what is typical for the diffuse ISM. It is not known presently whether 
all these events are related or have a very different origin.

Extreme scattering events (ESE) are sources that show dramatic decrease (often $>50$\%) in the flux density
at radio wavelengths, usually close to 1 GHz, and last for a period of several weeks to months.
The flux density changes observed in ESEs are very different from the flickering of radio sources
frequently observed at GHz frequencies and attributed to interstellar scintillation.
While scintillation typically causes persistent variability on 
timescales measured in days and at a level of few percent (although larger flux changes
are observed for ``intra-hour" variability, which is thought to be produced by nearby scattering 
with systematic 
increasing wavelength), variability observed in ESEs is much larger (often $>50$\%), 
smooth, lasts over several months,
and does not have usually strong frequency dependence 
except near the wavelengths at which rays converge, forming caustics.

In addition, the flux density variability of radio sources has also been observed 
at a level of few to 10\% and on timescales of 1-2 days. This was first noticed by 
\citet{Heeschen84} and \citet{Witzel86}, and is known as 
 intra-day variability (IDV). Since early days it was hypothesized that IDVs are  
caused by slow scintillation in the ISM and 
intergalactic medium.
As summarized by \citet{Jauncey16} and \citet{Bignall15},
the combination of time delay measurements, the discovery of $\sim$ 10 IDV sources that show an 
annual cycle in their variability,
as well as a strong correlation with Galactic latitude;
all overwhelmingly support the idea that  
interstellar scintillation is the primary cause of variability in IDVs. However, we emphasize
that some events defy simple classification (e.g. \citet{Tuntsov17}). 
As mentioned above, the
spatial distribution and properties of the scintillation material are still not fully
understood. Studies such as \cite{Cordes85} and \cite{Lazio08} suggest that scintillation results
from localized regions, possibly even clumps, distributed throughout the Galactic disk. Given
the many possible forms and distributions of interstellar density variations, and the possibly complex structure of the sources themselves, many forms of variation with time and frequency are evidently possible. As IDV sources have 
less extreme properties than ESEs and no,
to an order of magnitude,
overpressure problem \citep{Heiles07,Tuntsov13},
we do not include them in this review.

Pulsar dynamic spectra (frequency vs time) have a characteristic pattern due to interstellar scintillation 
and their Fourier transform is called the secondary spectrum.
Many sensitive observations have shown parabollic arcs in the secondary spectra, 
whose curvature depends on the location of the scattering material along
the line of sight \citep{Hill05,Stinebring07}. Arcs imply existence of many localized scattering screens along the line of sight, with
properties in accord with the Kolmogorov turbulence and $n_e$ typical for the ionized ISM.
However, some observations have shown small sub-structure within scintillation arcs, called ``arclets'',
which require small dense structures with a size of $\sim1$ AU, $n_e\sim 100$ cm$^{-3}$ 
and $P/k\sim 10^6$ K cm$^{-3}$ \citep{Hill05} to be explained.  

This review focuses only on the most extreme TSIS: ESEs which have extreme  over-pressure and over-density problems.
While properties of scintillation arclets 
appear similar
to those found in ESEs, and arclets have the over-pressure problem,
we do not include them in this review. 
We also note that interplanetary scintillation causes even faster, on scales 
of seconds, flux density variability 
of compact radio sources and is caused by inhomogeneities in the ionized solar wind.
A recent study of interplanetary scintillation using the Murchison Widefield Array (MWA) found evidence for structure on
spatial scales from $<1000$ to 10$^6$ km ($\sim100$ AU) \citep{Kaplan15}.

\subsection{Extreme Scattering Events}\label{s:ESE}

\begin{figure}
\includegraphics[scale=0.5,angle=-90]{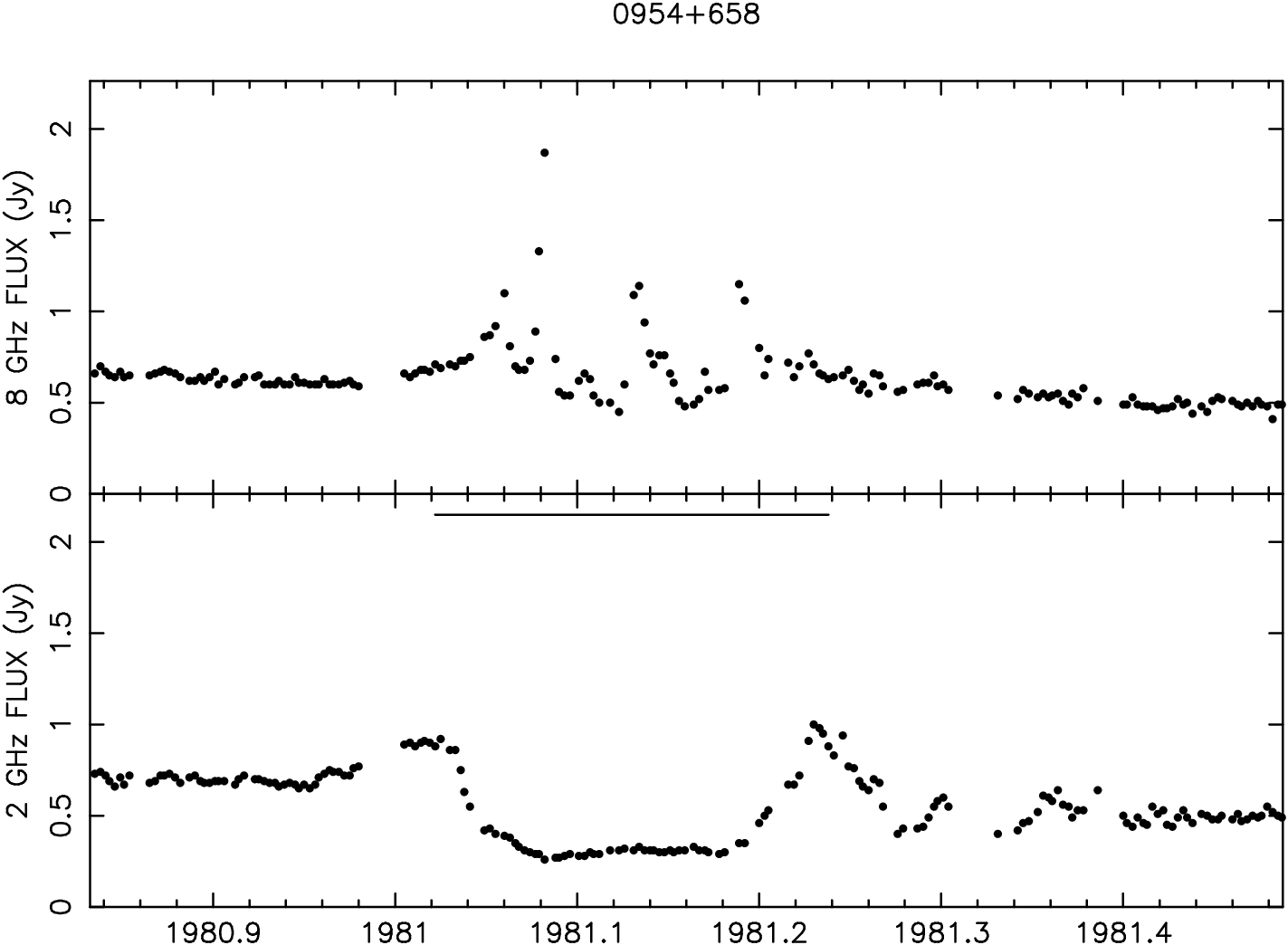}
\caption{The first discovered ESE. Flux density at 8 GHz (top) and 2.7 GHz (bottom) of QSO 0954+658 obtained with the Green Bank Interferometer. The x-axis shows time in years. The dramatic decrease of the flux density at 2.7 GHz around 1981.2 yrs, accompanied with several sharp spikes at 8 GHz, is caused by refractive effects due to QSO's light encountering a dense plasma lens. From Fiedler et al. 1987. }
\label{f:0954+658}
\end{figure}


\subsubsection{Early Studies}
The first ESE event was discovered by \citet{Fiedler87b}.
In their daily monitoring of 36 extragalactic sources using the Green Bank Interferometer 
over 7 years at 2.7 and 8.1 GHz, Fiedler et al. noticed unusual
minima in the light curves of 7 sources. QSO 0954+658 (see Figure~\ref{f:0954+658}) showed 
exceptionally dramatic light curves with a minimum at 2.7 GHz being correlated with several 
sharp spikes at 8.1 GHz. Based on the coordinated behavior at the two frequencies without any time lags,
 intrinsic source variability was excluded as a possible cause of variability and it was concluded that
variations are caused by an occultation event. The light travel time arguments suggested that
the occulter is in the Milky Way with a transverse velocity $< 300$ km/sec. Combined with the 
estimated proper motion from the rapid decline of the light curve, 
a distance of $<1.3$ kpc was implied.
From the duration of the light curve minimum an estimate of the size of the occulter was
provided, $<7$ AU. 
Trends at two frequencies suggested refractive effects through irregularities in the ionized gas density
of the occulter. Assuming a spherical lens, the implied density of the occulter was $n_e \sim 4 \times 10^4$ cm$^{-3}$, much greater than that of the 
diffuse ionized ISM. Fiedler et al. (1987) concluded that the occulting sources likely represent a new 
astrophysical phenomenon whose nature and stability need to be understood.

\citet{Romani87} provided a more detailed interpretation of the 0954+658 event and suggested that
the high pressure in the refracting structures is naturally found inside old SNRs 
(or the edge of a Galactic outflow), proposing
that SNRs could be natural sites for the plasma lensing sheets with the
over-pressure lenses residing inside the high-pressure environment. 
They made an estimate that on average 
a line of sight will pass through about 100 face-on sheets. This will cause small scattering angles, however if the sheets
have small density corrugations this can significantly increase the scattering angle. 

\citet{Fiedler94} summarized results from a monitoring of 40-150 sources over a period of 11 years with the
Green Bank Interferometer, and 330 sources over a period of six months using the NRAO 300-foot radio telescope. 
Only 10 ESEs were detected in the 2.7 GHz light curves. While the original ESE in the direction of 0954+658 had
counterparts at 8.1 GHz, most other ESEs  only showed changes in flux density at 2.7 GHz.
Also, a greater variety of light curve shapes has been discovered. 
\citet{Fiedler94} also plotted positions of their 365 observed sources, as well as nine sources
exhibiting one or more ESEs, and positions of four major Galactic radio continuum loops.
The ESE events appeared to be in a statistically significant 
association with Loops I, II and III.

While Fiedler et al. (1994) provided a statistical model for flux redistribution based on stochastic broadening
regardless of its nature (refractive or diffractive), \citet{Clegg98} 
provided a detailed quantitative 
description of the optical properties of an interstellar plasma lens assuming that this is a 
discrete object (Figure~\ref{f:clegg_schematic}). This model is still being used to
estimate lens physical properties from observed light curves. 
The 1-D refractive properties of the lens are specified by a dimensionless parameter $\alpha$:
\begin{equation}\label{eq:alpha}
\alpha = 3.6 \left ( \frac{\lambda}{1~cm} \right )^2 \left ( \frac{N_0}{1~cm^{-2}~pc} \right ) \left 
( \frac{D}{1~kpc} \right ) 
\left ( \frac{a} {1~AU} \right )^{-2}
\end{equation}
where $\lambda$, $N_0$, $D$ and $a$ are the wavelength of observation, maximum free-electron column density, 
distance from lens to observer, and size of the lens transverse to the line of sight, respectively. In the central portions of the lens, the overdensity of the plasma causes rays to
spread (diverging lens) corresponding in the light curve to reduce intensity. Rays passing through the wings of the lens, where the density gradient is largest, are differentially refracted, leading to formation of converged rays, or caustics, which correspond in the light curve to spikes. The formation of caustics may be sensitive to the density profile, e.g a power law would behave differently from a Gaussian. If the lens were aspherical, the light curve would be sensitive to viewing angle. While these refinements are possible topics for the future, the
Clegg et al. model has been remarkably successful in demonstrating the physical effects in
play and reproducing many of the observed features of ESEs.

A second
parameter, $\beta_s$, characterizes the extent to which the effect of the lens is diminished 
by the intrinsic size of the source:
\begin{equation}
\beta_s=\theta_s / \theta_l
\end{equation}
where $\theta_s$ and $\theta_l$ are the angular sizes of the background source and the lens, respectively. 
For $1/\beta_s >>1$ (lens much larger than the source) or $1/ \beta_s <<1$ 
(lens much smaller than the source), the expected amplitude variations are very small and are limited by 
the sensitivity of observations. For  $\beta_s$  of order unity, the behavior is more interesting, 
with deep flat minima and shallow rounded minima both possible.
Yet, sources may have multiple components and complex morphology, the effects of which are not easily encapsulated by a single parameter,
as perusal of Clegg et al.'s paper reveals.

By applying the model to light curves of two ESEs Clegg et al. estimated lens sizes of 0.065
and 0.38 AU and electron densities of
$n_e=300$ and 10$^{5}$ cm$^{-3}$, respectively. 
The implied thermal pressures of $P/k\sim 10^{6-9}$ K cm$^{-3}$ are very high and Clegg et al. concluded that such discrete, highly over-pressured lenses are either highly transient or are embedded in high-pressure environments. 
Both ESEs (0954+658 and 1741-038) were found to have likely associations with radio Loops III and I.

Besides explaining the minimum in the background source's light curve due to the presence of a diverging plasma lens,
this model also predicted several additional observational effects: (1) the formation of caustic surfaces
which have been observed in many ESEs (e.g. 0954+658 in Figure~\ref{f:0954+658}
shows caustic spikes at the start and end of the ESE event),
(2) possible creation of multiple source images (discovered later in the case of 2023+335, Pushkarev et al. 2013), (3) 
angular position wander of the background source.
While all these effects have been observed in ESEs supporting the discrete-lens model, some discrepancies 
between the model and observed light curves have 
been noticed since early days. Clegg et al. (1998) suggested that the discrepancies could be due to some
combination of sub-structure within the lens, an anisotropic lens shape, a lens that
only grazes the source instead of passing completely over it, or possibly unresolved sub-structure within the extragalactic source.

\subsubsection{More Recent Results}

\citet{Lazio00} obtained VLBI imaging of PKS 1741-038 at four epochs during and after an ESE event. While multiple source 
images and the angular position wander were not detected, the source exhibited excess angular broadening during the ESE phase. However, the amount of angular broadening was larger than what is expected by simple 
refractive defocussing (Clegg et al. 1998).
In addition, these four-epoch measurements hinted at the existence of an anti-correlation between the source flux density and angular diameter. Lazio et al. suggested a likely combination of refractive defocussing by a discrete lens and stochastic broadening caused by interstellar turbulence.
This experiment shows a power of frequent monitoring of ESE sources. 
The correlations between the source flux density and angular broadening clearly needs further observations and testing.

As pointed by Fiedler et al. (1994) and many others, one of the key open questions about ESEs is their relationship
with other phases of the ISM, specifically TSAS. \citet{Lazio01a} is the only study of HI absorption of
a source (PSR 1741-038) while it was undergoing an ESE. 
No changes in the HI absorption spectra, or rotation measure, were detected. Their upper limit on the HI optical depth was $\Delta \tau<0.05$, marginally ruling out the existence of TSAS associated with this ESE. As this is the only such experiment, additional and even more sensitive measurements are needed to investigate possible connection between TSAS and ESEs.


\citet{Pushkarev13} detected for the first time significant structural changes in QSO 2023+335 images over time,
as well as changes in the light curve typical of an ESE at 15 GHz. 
The observations were obtained over 16 epochs (over 11 years) and similar structural changes were not found in $\sim300$  other AGN jet sources observed by the Monitoring Of Jets in Active galactic nuclei with VLBA Experiments (MOJAVE) program. The observed structural changes are due to
multiple imaging of the source induced by refraction in the intervening ISM. 
The angular separation between image peaks follows a $\lambda^2$ dependence, in agreement with a plasma scattering origin of the induced sub-images.
There are several additional unusual aspects of this source. First, while most ESEs are detected at 2 GHz, this is the only one detected at 15 GHz 
(as well as 8 GHz). Second, multiple ESE events have been noticed in the light curve of this source. 

By fitting the light curve with the stochastic model by Fiedler et al. (1994), Pushkarev et al. estimated the  
apparent angular size of the lens ($\theta_l=0.27$ mas). 
As QSO 2023+335 is located behind the Cygnus supernova remnant, it was assumed that the lens is associated with the Cygnus region and is at a distance $D=1.5$ kpc. The 
transverse linear size of the lens is: $a=\theta_l D=0.4$ AU. 
The proper motion of the lens of 6.8 mas yr$^{-1}$ results in the transverse lens speed of 48.7 km/sec.
Using these estimates and equation (1), they obtained $N_0=2.5 \times 10^{17}$ cm$^{-2}$
and a free-electron density within the lens $n_e=N_0/a= 4 \times 10^4$ cm$^{-3}$.
Interestingly, their analysis of the angular broadening suggested an unusually steep turbulent spectrum with a slope 4.2-4.7. This suggests that the over-dense and over-pressured lens responsible for the observed ESE lives inside a scattering medium
with an unusually steep turbulent spectrum. While the origin of such turbulence is still not clear, more shallow spectra 
were seen in previous studies and were associated with localized (instead of uniform) scattering screens.

\citet{Bannister16} argued that real-time detection is the key to understanding  ESEs
as most properties of the lens are measurable only while ESE is in progress (e.g. $n_e$ profile,
angular geometry of the lens, dust/neutral/magnetic field properties of the lens).
A dedicated new, real-time search for ESEs is underway with the Australia Telescope Compact Array. About 1000 AGNs are being monitored once a month, with only 50 sec of integration time on target over the  4-8 GHz frequency range. An automatic search for changes in the continuum spectrum from
a smooth distribution are used  to 
identify new ESEs. Dedicated high-cadence monitoring is then 
performed for identified events.
An exciting result from this campaign is the discovery of 
an ESE in the direction of PKS 1939-315 in 2014.
The detailed monitoring allowed this study to model the $N_e$ profile as a function of time. They found that the peak in the profile corresponds to the minimum of the flux density profile, suggesting a diverging lens.
The column density changes of $N_e \sim 10^{16}$ cm$^{-2}$ over transverse scales of 
$\sim10^{13}$ cm (corresponding to 20 days at an assume transverse velocity of 50 km/s), implied $n_e \sim 10^3$  cm$^{-3}$ and $P/k \sim 10^6$ K cm$^{-3}$ (if no elongation along LOS assumed). 
The goal of this program is to generate a statistically significant sample in the future to determine covering fraction and spatial distribution of ESE lenses.


\newpage
\begin{table}[h]
\caption{Summary of ESE events.}
\label{t:ESEs}
\begin{center}
\begin{tabular}{@{}l|c|c|c|c||c@{}}
\hline
Source & Coords ($^{\circ}$)& Size (AU) & $n_e$ (cm$^{-3}$) & $N$ (cm$^{-2}$) &Reference\\
\hline
0954+658 &146,43&0.4 &$10^5$ &$6\times 10^{17}$ & Clegg et al. 1998\\
1741-038 &21.6,13.1&0.065 &300. &$3 \times 10^{14}$ & Clegg et al. 1998\\
0133+476 &&-& -& -&  Fiedler et al. 1994\\
0300+470 &&-& -& -&  Fiedler et al. 1994\\
0333+321 &159.0,$-18.8$&-& -& -&  Fiedler et al. 1994\\
1502+106 &11.4,54.6&-& -& -&  Fiedler et al. 1994\\
1611+343 &&-& -& -&  Fiedler et al. 1994\\
1749+096 &21.6,13.1&-& -& -&  Fiedler et al. 1994\\
1821+107 &&-& -& -&  Fiedler et al. 1994\\
2352+495 &&-& -& -&  Fiedler et al. 1994\\
2023+335 &&0.4&$4 \times 10^4$ &$2.4 \times 10^{17}$ & Pushkarev et al. 2013\\
PKS1939-315 &8.5,$-24.0$&0.7 &$10^3$ &$10^{16}$ & Bannister et al. 2017\\
PSR J1643-1224&5.7,21.2& 56. & 130. & $10^{17}$ & Maitia et al. 03\\
PSR B1937+21 &57.5,$-0.3$& 0.094& 25. & $3\times 10^{13}$& Cognard et al. 93\\
PSR B1937+21 &57.5,$-0.3$&0.05 & 220.& $2\times 10^{14}$&  Cognard et al. 93\\
PSR B1937+21 &57.5,$-0.3$& 0.6 & 200.& $2\times 10^{15}$ &Lestrade et al. 98\\
PSR B1800-21&8.4,0.1& 220.& 15000.& $5\times 10^{19}$& Basu et al. 2016\\
PSR J1603-7202&316.6,$-14.5$& 4.9& 3.4& $2\times 10^{14}$& Coles et al. 15\\
PSR J1017-7156&291,$-12.6$& 13.9& 3.7& $8\times 10^{14}$& Coles et al. 15\\
0133+476& 131,$-14$&-& -& -& Lazio et al. 2001\\
0201+113&150,$-47$&-& -& -&  Lazio et al. 2001\\
0202+319&141,$-28$&-& -& -& Lazio et al. 2001\\
0300+470&145,$-10$&-& -& -& Lazio et al. 2001\\
0528+134&191,$-11$&-& -& -& Lazio et al. 2001\\
0952+179&216,48&-& -& -& Lazio et al. 2001\\
0954+658&146,43&-& -& -& Lazio et al. 2001\\
1438+385&66,65&-& -& -& Lazio et al. 2001\\
1502+106&11,55&-& -& -& Lazio et al. 2001\\
1756+237&49,22&-& -& -&  Lazio et al. 2001\\
2251+244&92,$-31$&-& -& -& Lazio et al. 2001\\
2352+495&114,$-12$&-& -& -&  Lazio et al. 2001\\

\hline
\end{tabular}
\end{center}
\begin{tabnote}
Note that nine ESEs were discovered by Fiedler et al. (1994) but do not have estimated size and density of the intervening lenses. ESEs detected in Lazio et al. (2001) by applying a wavelet analysis on the Green Bank Interferometer monitoring data also do not have estimated size or density.
\end{tabnote}
\end{table}

\newpage
\subsubsection{Pulsar ESEs}

Similar dramatic changes in the flux density have been seen in the case of several pulsars. \citet{Cognard93} observed a sudden decrease in the flux density of PSR1937+21 over a period of 15 days,
coupled with an increase in the time of arrival of the pulses from this millisecound pulsar. These observations are modeled as resulting from perturbations in the dispersive delay 
(caused by DM fluctuations which are proportional to fluctuations in the electron density), and the geometric refractive delay (caused by the intervening ionized screen and being proportional to the refraction angle and the distance to the screen). The best fit was obtained for a two-component model with
electron densities of 25 and 220 cm$^{-3}$, respectively, and corresponding
transverse structure lengths of 0.094 and 0.05 AUs.  
While the inferred densities are lower than what is typically found for ESEs, they still imply a large over-pressure problem.
PSR1937+21 is in close angular proximity to the Cygnus loop and Cognard et al.
concluded that these observations support models for ESE production near peripheries of expanding superbubbles.
Finally, they suggested that pulsar ESEs could be the dominant source of timing noise in pulsar observations and need to be considered when
searching for signatures of gravitational waves in pulsar timing measurements.
\citet{Lestrade98} analyzed observations of the same pulsar over an even longer period and identified several additional ESEs.
For the Oct 1989 event (same as in Cognard et al. 1993) they estimated approximate transverse size of 0.6 AU assuming the velocity 
of 50 km/s, and density of 200 cm$^{-3}$.
Their work showed how difficult it is to identify ESEs. 

\citet{Basu16} studied PSR B1800-21 at several frequencies using the GMRT. This is a young pulsar within the W30 complex which contains a SNR and a number of HII regions. This pulsar, and a handful of others, exhibited a significant evolution during the observing period with the low-frequency part
of the spectrum becoming steeper for a period of several years before reverting back to its initial shape.
The most likely reason for the turnover is free-free absorption by the intervening medium, whose properties have been estimated as having a size of 220 AU and $n_e\sim1.5 \times 10^4$ cm$^{-3}$. While this was not a classical pulsar ESE event, the observed properties are very much in the ESE range.

\citet{Coles15} showed two pulsar ESE events with the corresponding dispersion observations. The ESEs were seen at the same time in DM fluctuations ($\delta$DM), the coherence bandwidth ($\nu_0$) and the intensity timescale ($\tau_0$), and were identified by eye. The estimate size is about 5-14 AU and $n_e \sim4$ cm$^{-3}$. The DM fluctuations were 0.0023 and 0.0015 pc cm$^{-3}$, respectively. However, in addition they found several cases of significant DM fluctuations that did not have the corresponding changes in the coherence bandwidth or time fluctuations and therefore would not be identified as ESE. We summarize basic observed properties of ESEs in Table~\ref{t:ESEs}.

\begin{figure}
\caption{A schematic diagram of refraction by a Gaussian plasma lens from Clegg et al. 1998. The top panel shows ray paths. Rays which pass near the cloud center diverge, while ray paths which pass through the sloping edges are differentially refected and converge, forming caustics. }
\label{f:clegg_schematic}
\end{figure}

\subsection{Theory of Ionized Structures}
\label{s:ionized_theory}

\subsubsection{Refraction Basics}
\label{s:refraction-basics}

The theory of scintillation - both wave propagation and how it translates to observables - is well documented elsewhere \citep{Rickett77,Rickett90}, so we only quote a few basic results here. The refractive index $n_r$ of electromagnetic waves in a plasma is:
\begin{equation}\label{eq:nr}
n_r=\left(1-\frac{\nu_{pe}^2}{\nu^2}\right)^{1/2}\sim 1 - \frac{\nu_{pe}^2}{2\nu^2},
\end{equation}
where $\nu_{pe}\equiv (n_ee^2/\pi m_e)^{1/2}= 9.0\times 10^3n_e^{1/2}$s$^{-1}$ is the electron plasma frequency and the second relation holds for $\nu_{pe}/\nu\ll 1$, which generally holds for observations of the ISM. 

According to eqn. (\ref{eq:nr}), radio frequency waves have group velocity $v_g\sim c(1-\nu_{pe}^2/2\nu^2)$. Pulsar pulse arrival times $t_p$ 
therefore depend on $\nu$; 
 $dt_p/d\nu\propto\nu^{-3}$. This can be used to derive the total electron column density, or dispersion measure DM, between the
pulsar and Earth.
The importance of this goes beyond its value as a probe of the ISM. Pulsar pulse arrival times are sensitive to low frequency gravitational waves in the spacetime between the pulsar and the Earth \citep{vanhaasterin2014}. The dispersion in pulse arrival times caused by the theoretically predicted gravitational wave background is typically tens to hundreds of nanoseconds, while
temporal variations in DM cause dispersion of a few $\mu$sec.
This creates a huge incentive to monitor and model temporal variations of DM and all 
other plasma effects so they can be properly
accounted for in analyses of pulsar timing. Serendipitously,  pulsar timing
data can be used to monitor electron density structure in the ISM \citep{Lam2016}. 

Variation of $n_e$ with position on scales large compared to a wavelength causes refraction. Equation (\ref{eq:nr}) can be used to explain the divergence of rays shown in Figure 9. The waves are refracted toward regions of lower phase speed, which are regions of lower electron density
 (see eqn. (\ref{eq:alpha})).
Variations of $n_e$ also cause variation of the source intensity, angular broadening and, in the case of pulsars, temporal broadening. It can be shown from the equations of geometrical optics that a wave of wavelength $\lambda$ 
which
propagates a distance $L$ through density fluctuations with transverse scale $l_{\perp}$ and amplitude 
$\Delta n$ is scattered through an angle $\Delta\theta_0$ of order $r_e\lambda^2(L/l_{\perp})\Delta n$, where $r_e\equiv e^2/m_ec^2$ 
is the classical electron radius. Over the distance $D$ to the source the time of flight of a ray deflected into
our line of sight is increased by $\Delta\tau\sim D(\Delta\theta_0)^2/c$. If there are $N\propto D$ deflections and the $\Delta\theta$ have a Gaussian distribution with rms value $\Delta\theta_0$, then the temporal broadening of pulses should scale with distance, density fluctuation
amplitude, and wavelengths as $D^2(\Delta n)^2\lambda^4$.

\subsubsection{Nature of the fluctuations}\label{s:nature}
Soon after the discovery of pulsar scintillation \citep{Rickett70}, it was pointed out by \citet{Scheuer70} 
that the models in which the density fluctuations are plasma waves require 
special conditions and large energy inputs. \citet{Scheuer70} estimated the power requirements to be about $1.6\times
 10^{-24}$ erg cm$^3$ s$^{-1}$, which is almost an order of magnitude larger than the energy 
input from supernova explosions and 2-3 orders 
of magnitude higher than the power supernovae supply to the kinetic energy of the ISM.
Although one may quibble with the characterizations of the fluctuation spectrum and its 
dissipation mechanisms in \citet{Scheuer70}, their basic points  that ion-neutral 
damping is strong in partially ionized regions, and that wave dissipation is strong at these short wavelengths, 
remain valid and impose strong constraints on any theory for interstellar scintillation.
 
Motivated in part by the high density and ionization fractions inferred for the 
scattering environment, attention turned to ionized portions 
of the ISM as the source of scintillation.
The collisionally ionized gas in the ISM (HIM) is probably too low in density to be the site of TSIS, but warm, photoionized  
gas (the WIM) has a moderately large filling factor and typically density $n_H \sim 0.05 - 0.1$ cm$^{-3}$. We distinguish the
WIM from H II regions around hot stars, which can be quite dense and overpressured, 
and from the edges of clouds, which can also be overpressured if the clouds are 
self gravitating. After ionized gas was identified as the source of scintillation, the question of whether the host medium was confined or pervasive remained. However, large density enhancements, and therefore large densities, were favored.

Higdon \citep{Higdon84,Higdon86} proposed an ingenious solution to the energetics problem. 
In his picture, the density fluctuations arise from essentially static entropy modes, 
which are nearly isobaric ($\delta n/n\sim - \delta T/T$) and cascaded from large to small scales due 
to interacting with a turbulent, highly anisotropic spectrum of MHD waves
propagating almost perpendicular to the background magnetic field. 
In this scenario,  $n_e$ is merely a passive scalar. Higdon argued that such 
waves are relatively weakly damped, and that because they are driven by a cascade 
originating at large scales, they tap into the global energy reservoirs in the 
ISM. Although Higdon considered a number of physical settings, he ultimately  
favored turbulence driven by evaporating clouds in HII regions surrounding hot stars, 
which drive both thermal and dynamical perturbations to the surrounding medium,
as the sites where scintillation is produced \citep{Higdon86}.

Higdon's ideas were extended and  made more rigorous  in a series of papers on interstellar 
turbulence beginning with \citet{Goldreich95} and
 culminating with \citet{Lithwick01}. These papers developed the theory of anisotropic, 
turbulent cascades of shear, compressionless Alfven waves. The anisotropy scaling  within the inertial
range of  the Goldreich-Sridhar spectrum can be derived by assuming constant energy 
flux in $k$ space ($k_{\perp}v_{\perp}^3(k_{\perp})\sim k
_0v_A^3$)  and that the turbulence is
 critically balanced ($k_{\perp} v_{\perp} (k_{\perp}) \sim k_{\parallel} (k_{\perp}) v_A$), 
where $k_0$ is the scale at which the turbulence becomes Alfvenic. 
As we noted in \S 1.3,
these two relations lead to elongated contours of constant power in $k$-space with $k_{\parallel}/k_{\perp}\sim (k_d/k_{\perp})^{1/3}$.
Strictly this 
only means that contours of constant power in the ($k_{\perp}, k_{\parallel})$ plane are elongated, but coherent structures tend to be aligned with the magnetic field as well. 
 
When weak compressibility effects are included, the turbulence can be resolved into three 
wave modes: the shear Alfven mode, which is compression free,
and the fast and slow magnetosonic  modes. While both magnetosonic modes 
produce density fluctuations, the slow mode is more compressive and has a spectrum similar to the Kolmogorov spectrum
\citep{Cho02}, in agreement with the scintillation measurements. 
Thus, there are two sources of density fluctuation in the \citet{Lithwick01} theory: entropy modes and slow modes. 


The slow
modes may be important. As Lithwick and Goldreich pointed out, when the cooling time of 
the entropy modes is short compared to the eddy turnover time at their scale, they are 
heavily damped, and therefore cannot be cascaded from large scales. And, the power spectra of passive 
scalars are in general not the same as that of the turbulent velocity fluctuations that mix 
them \citep{Shraiman00,Warhaft00}, but have exponential tails and a highly 
intermittent  spatial distribution. As of this writing, the electron density fluctuation spectrum produced by 
entropy and MHD modes under ISM conditions, i.e. with realistic damping, cooling, and recombination has 
not yet been definitely calculated. 
While the observed spectrum of density fluctuations is roughly in agreement with expectations for slow magnetosonic turbulent modes, it is still not understood why and how the observed spectrum persists to very small spatial scales where theoretically we would expect the entropy modes to be heavily damped.

\subsubsection{Distribution of scattering material}

The dependence of scattering on source
distance and location in the Galaxy provides important 
information that complements what
can be gleaned from the power spectrum. We showed earlier in this section that if the angular 
deflection of radio waves due to scattering follows Gaussian statistics, then the widths of pulsar pulses should scale with wavelength and 
distance as $\lambda^4D^2$. However, for pulsars 
with dispersion measures (or electron column densities) greater than about 20 cm$^{-3}$ pc, 
the relationship is better fit by  $\lambda^4D^4$. 
 This was taken as evidence by \citet{Cordes85} that the scintillation cannot be entirely 
due to a pervasive homogeneous medium but must include a contribution from
localized clumps within which enhanced scattering occurs. 

The two component model  was formalized and generalized in a series of papers by Boldyrev  and Gwinn.
In \citet{Boldyrev03,Boldyrev06} it is argued that the $D^4$ scaling of the pulse width with 
distance  can be explained if the distribution
 of $\Delta\theta$ is not Gaussian, but rather is described  by 
a  Levy distribution.  Levy distributions have a divergent second moment. Thus, quantities 
such as the mean squared scattering angle $(\Delta\theta)^2$ or pulse width $\tau$ are 
dominated by a few of the largest scattering events. Whereas in a Gaussian model the mean squared angular 
deviation scales linearly with $D$ and the time delay therefore scales as $D^2$, the observations 
suggest that the mean squared angular deviation scales as $D^3$. A $\Delta\theta$ distribution $P(\Delta\theta)
\propto  \vert\Delta\theta\vert^{-5/3}$ will lead to this dependence.

Remarkably, \citet{Boldyrev06} showed from geometric arguments that rays passing 
through thin shells of ionized gas are scattered according to just such a distribution. The ``rare events"
in this case correspond to ray paths that just graze the inner edges of the shells, resembling sheets seen edge on. They suggested that
the scattering  arises in the ionized edges of molecular clouds near luminous hot stars 
and present quantitative arguments for the properties of these shells: $n_e\approx 10^2$ cm$^{-3}$,
shell thickness of $\sim 0.003$ pc, cloud radii of about 10 pc, and cloud separations of 
about 100 pc. The ionized shells may themselves become turbulent, leading to additional scattering.

\subsubsection{Kinetic effects}

An entirely different physical explanation for a similar Levy distribution was proposed 
by \cite{Terry07,Terry08,Smith11}. These authors  studied decaying Alfven turbulence
 down to wavelengths of a few to tens of thermal ion gyroradii $r_i\equiv (k_BT_e/m_i)^{1/2}/\omega_{ci}$. In contrast to the shear Alfven waves in the magnetohydrodynamic cascade, which do not
 have density fluctuations, as the wavelength approaches $r_i$ the electron density begins to fluctuate in tandem with fluctuations
 in the parallel current such that:
%
\begin{equation}\label{eq:kaw}
\frac{\delta n_e}{n_e}\sim\frac{\delta B_{\perp}}{B}\frac{k_{\perp}r_i}{(1+k_{\perp}^2r_i^2)^{1/2}}.
\end{equation}
Thus, while at long wavelengths electron density is advected like a passive scalar, 
at the short lengthscales associated with pulsar scintillation, density fluctuations play a key role in the dynamics of the waves, which in this regime are known 
as kinetic Alfven waves. When nonlinear
effects are included,  electrons  form coherent sheets and filaments with density 
profiles that  scatter radio waves according to the same $\Delta\theta$ distribution as thin shells. 
To the extent that the shells are turbulent, with turbulence extending to kinetic scales,
 the two pictures are not mutually exclusive. 
 
 The question of what is the 
{\textit{minimum}} scale for kinetic Alfven wave turbulence is not fully resolved. 
At $k_{\perp}r_i\ge 10$, the waves are 
 collisionlessly damped by electrons, but \cite{Howes11} found that turbulent 
power persisted down to the electron gyroscale $r_e$, consistent
 with the range of scales observed in the solar wind. Evidence that the electron 
density fluctuations extend to scale as small as 70 km, as small as $r_r$, is 
presented in \cite{Rickett09} and reviewed in \citet{Haverkorn13}.

At this point, there is no clear consensus on the nature of the electron density fluctuations 
revealed by pulsar scintillation. It is probably safe to say that they occur in gas that is 
nearly fully ionized and denser than the hot, collisionally ionized component or warm diffuse 
component that is not specifically associated with bright stars. Whether  they are produced by 
an underlying turbulent process or are associated with particular types of object is less clear. 
When we consider that the pulse width-distance relation can in principle, be
 produced by scattering from thin shells of ionized gas 10 pc in size, or electron filaments and 
sheets 1000 meters in size, and that H II regions, cloud edges, and supernova shells can all host turbulence with Kolmogorov features, the possibilites seem wide indeed.

\subsubsection{ESEs}
\label{s:ESE-theory}

An important related question is whether ESEs are a more 
extreme but qualitatively similar form of scattering
 \citep{Hamidouche07} or are a signature of
something entirely different. The ESE discovery paper \citet{Fiedler87a} suggested that the 
transient brightness dip is due to refraction by an overdense ``plasma lens"  with projected size on the sky of a few AU. We reviewed the properties of plasma lenses in \S\ref{s:ESE}, and only note here that the recent work of \cite{Bannister16} is considered to rule out underdense, magnetically structured lenses
such as suggested by  \citet{Pen12}.
If such lenses are roughly spherical then $n_e\sim 10^3 - 10^4$ cm$^{-3}$, which together 
with typical ionized gas temperatures $T\sim 10^4$ K
 imply pressures at least 10$^3$ times higher than the typical ISM
pressure. It was immediately appreciated that without a confinement mechanism such 
structures could survive for at most a few years.

%
%

If the lenses are thin, turbulent sheets viewed end on they could have high column density 
without high volume density, thus reducing the overpressure problem. For example, 
if the elongated structures followed Goldreich-Sridhar  scaling, the mean 
density of a structure viewed end on would be $(l_{\perp}/l_{\parallel})^{2/9}$ smaller 
than if the structure were spherical; for $l_{\perp} = 10^{13}$ cm, $l_{\parallel}\sim 1$ pc, 
this is a factor of  .06. However, because of the small transverse dimension, extremely 
thin substructure is still required. In the model of \citet{Romani87}, the extreme brightening 
arcs are due to caustics, or converging ray paths; in the turbulent walls of old supernova remnants seen edge on. 

%
%
%
%

\cite{Walker98} and \cite{Wardle99} attempted to solve the high pressure problem for ESEs by suggesting that the  
refraction occurs  in ionized evaporative outflows from  cold, self gravitating  clouds a few 
AU in size.  If the clouds have  velocity dispersions characteristic of halo objects, their motion through all but the hottest portions of the ISM drives bowshocks, and creates a 
hot gas layer between the shock and the cloud which intensifies the evaporation, and possibly 
draws out a turbulent magnetotail. All of these phenomena create dense, small scale structure which 
may explain a variety of ESE light curves and pulsar scattering properties. Based on a covering factor $f_c\sim.005$ \citep{Fiedler94}, the tiny clouds postulated 
by Walker \& Wardle are a major form of dark matter in the Milky Way (optical lensing by
similar clouds was proposed by \cite{Draine98} as the source of microlensing in dark matter surveys). However, 
as noted in \S~\ref{s:ESE_summary} statistics are still poor,
so it is possible that $f_c$ is lower and the clouds are more rare. The ongoing monitoring project for ESEs will place valuable constrains on  $f_c$. More fundamentally, \cite{McKee01} showed that self gravitating, polytropic clouds with the low surface pressures and high total column densities proposed for the ESE models cannot be heated sufficiently
by standard mechanisms to offset radiative cooling. While it may be possible to create
composite models with a polytropic index that varies from core to envelope by fine tuning, but this throws up a serious roadblock to self gravitating clouds as a source for ESEs. 


In our view, a turbulent origin of ESEs cannot be ruled out, but it probably requires a special environment in which driving is so strong that the cascade reaches kinetic scales at which electrons and ions decouple. Although ray tracing has been carried out for the plasma lenses reviewed in \S 3.1.1, we are not aware of any comparable study for electron
turbulence. Since simulations of such turbulence are now available, this is a viable project for the future.

\subsection{Summary and Outstanding Questions}
\label{s:ESE_summary}


{\bf Abundance.} In total, no more than 35 ESEs have been discovered so far (Pushkarev et al. 2013).
We list in Table~\ref{t:ESEs} their observed properties from the literature. The first one is the ESE from the discovery paper by Fiedler et al. (1987) that was studied by several authors.
Fiedler et al. (1994) discovered 9 additional ESEs, however they did not derive lens size and electron density. We list these sources in the table for completeness and potential future followups.
With over 300 sources observed and only 10 ESEs detected in over 11 years, Fiedler et al. (1994) estimated the rate of ESE occurrence as $\sim0.017$ events per source-year.
\citet{Lazio01b} performed probably the most systematic search so far. They monitored
149 sources with the Green Bank Interferometer every 2 days spanning 17 years and 
applied a wavelet-based technique to identify ESEs from the source light curves.
They discovered 15 ESE (lens size and density were not provided), however their wavelet technique failed to pick up several previously identified ESEs. In Pushkarev et al.'s MOJAVE sample of $\sim300$ AGNs only one ESE was found.
The current monitoring program by Bannister et al. is targeting 1000 AGNs and at least three ESEs have been discovered so far \cite{Bannister16,Bannister16ATel}.

While the statistical samples are small and the frequency of ESE events is still not properly established, it is clear that these are rare events. Many authors have pointed out, however, that
identification of ESEs is very difficult. For example, short-timescale and small-amplitude ESE
can be confused with interstellar scintillation, while long-timescale and small-amplitude ESE can be obscured by an intrinsic source variability. The exact definitions of how to 
identify and classify small- from high-amplitude variability is tricky.
At the same time, the connection between discrete lenses (commonly used in ESE interpretation) and the
turbulent ionized medium the signal from background sources is traveling through is still not clear.
One reasonable assumption could be that event timescales and amplitudes fall
in a continuous range that merge both scintillation and intrinsic variability.  
In addition, most ESEs have been discovered serendipitously. As ESE events can start at any time, it is unclear whether the cadence of observational experiments has been good enough in previous experiments. Frequent and dedicated monitoring, like in Bannister et al., is needed to provide proper statistical samples and estimates of the volume filling factor of ESEs.

ESE identification methods so far have been rather simplistic.
Most studies have identified ESE from light curves by eye. 
With larger monitoring programs systematic and automatic ESE searches are essential.
Lazio, Waltman, Ghigo et al. (2001) tested several statistical tools, in particular the wavelet 
transform of the light curves.
It is encouraging that they found 15 events in the light curves of 12 sources, however 
five ESEs previously identified by hand did not pass their wavelet selection criteria. This clearly requires further attention.
Bannister et al. are taking a new approach by monitoring AGN continuum spectra once per month over the 4-8 GHz frequency range, searching for sudden changes in the flux density 
{\textit{vs}} frequency spectra. As the plasma refractive index depends on $\lambda^2$, large departures from the power-law flux density spectra are expected at the start of an ESE event. 
In addition, as suggested by Clegg et al. (1998),
followup imaging observations right from the start of an ESE can provide information about
angular displacements and multiple images which can assist in the interpretation of light curves.
This requires very high resolution monitoring (e.g. with VLBI) as changes can occur on milli-arcsecond, or even smaller, angular scales. 

{\bf Likely sites of ESEs.}
Since early days a possible connection between SNRs and ESEs has been proposed. Fiedler et al. (1994) noticed angular proximity of nine ESEs with Loops I, II and III. ESE in the direction of QSO 2023+335 is positioned behind the Cygnus SNR. The ESE in the direction of PSR B1937+21 is also likely associated with a SNR shell.
We view detailed modeling of turbulence in and surrounding supernova remnants as a promising line of future research. Beyond the works of \cite{Romani87,Boldyrev06,Lam2016}, our understanding of turbulence associated with supernova driven shocks is growing rapidly due to its serendipitous connection with cosmic ray acceleration and transport \citep{Amato14}. This includes study of the interaction between the shock and pre-existing density irregularities, which
generates turbulence both downstream \citep{Beresnyak09} and upstream \citep{Drury12} of the shock. While these and other modelers have not had ESEs or other electron density fluctuations in mind, their work could be a springboard for a new study of small scale structure associated with supernova remnants. 

{\bf Internal structure.} Many studies have pointed out that understanding of the internal, and possibly multi-phase nature, of ESE lenses is still unconstrained. Only one ESE was followed up with a HI absorption experiment. While this experiment marginally excluded the possibility of TSAS existing along the same line of sight during the duration of the ESE, more sensitive observations are clearly needed. Additional measurements are necessary to investigate a possible TSAS-ESE connection.
Draine (1998) suggested that if ESE lenses have a molecular component there should be detectable H$_2$ vibrational absorption lines. James Webb Space Telescope (JWST) would provide an excellent possibility to search for H$_2$ associated with ESEs. Detecting these transitions during a lensing event would confirm the gaseous nature and determine the lens radial velocity, providing an ``unambiguous signature of gaseous lensing''. 
In addition, for IDV and intra-hour variable sources with nearby scattering screens, a deep search for the associated molecular gas with ALMA, as well as deep H$\alpha$ imaging to search for the ionized component, may be possible (Hayley Bignall, private communication).

Furthermore, detecting DM variations due to an ionized TSAS skin are now possible due to
extremely high-precision pulsar timing for gravitational wave detection.
Lam et al. (2016) estimated that the DM increase due to a pulsar passing behind
an ionized structure of electron density $n_e$ is: $10^{-5} n_e l_{10~AU}$ pc cm$^{-3}$, where
$l_{10~AU}$ is the pathlength through the structure. For a TSAS of 100 AU in size
and density of $10^3$ cm$^{-3}$, this would suggest $10^{-5} \times 10^3 \times 
10 \times 10^{-2}=10^{-3}$ pc cm$^{-3}$, assuming that only 1\% of the TSAS diameter is in the ionized skin.
Similarly, as pulsars move through the neutral medium they can ionize their surroundings and form bow shocks
essentially making their own scattering medium (as distinct from their magnetospheres).
Considering typical pulsar parameters, Lam et al. (2016) estimated the bow 
shock scales of AU to 0.1 pc and the resultant increase in DM 
increase at a level of $ 3.3 \times 10^{-3}$ pc cm$^{-3}$. 
This is typically of order 10$^{-3}$ of the mean DM.

{\bf The formation mechanism:  discrete lenses or fluctuations in the turbulent cascade?}
As we have shown so far, many parallel questions concerning the origin of TSAS and ESE still remain. The issue of whether ESE properties can be explained as being due to the turbulent cascade in the ionized medium, as opposed to discrete blobs or lens, is one of those questions. 
Detection of multiple images in particular lends support to the discrete lens model, yet only one or two such cases have been observed.
On the other hand, properties of most pulsar ESEs appear to be more mild, in terms of size and density, than those of ESEs discovered in the direction of extragalactic sources. This could be real, or due to some non-understood systematic differences in observing methods.
In fact, Hamidouche \& Lestrade (2007) argued using simple simulations that pulsar ESEs can be explained with Kolmogorov turbulence. 

The famous big power-law by Armstrong et al. (1995) was compiled from pulsar and extragalactic observations. For the finite-size extragalactic observations, Armstrong et al. (1995) used the
observed refractive index $n_r$ and source angular size (with an assumed distance to the scintillating medium of 500 pc) 
to estimate the amplitude of the power spectrum of electron density fluctuations, $C_n^2$, by using equation (17) from Coles et al. (1987). The power spectrum of electron density fluctuations is defined as:
\begin{equation}
P_{\delta n_e} (q) = C_n^2 q^{-\alpha},
\end{equation}
where $q=1/l$, with $l$ being the size or spatial scale of fluctuations.
Under the assumption that all sources trace the same ionized medium, a Kolmogorov slope of $11/3$ was used to estimate $P_{\delta n_e} (q)$ from the above equation. As 
a single power-law function can be fitted through all measurements of $P_{\delta n_e} (q)$, this is interpreted as electron density fluctuations in the ISM following a universal turbulent cascade. 

As an illustrative example only we asked the question of where would ESEs toward extragalactic sources land in this Big Power Law graph. The Pushkarev et al. ESE in the direction of QSO 2023+335 is a clear example of strong scattering (due to the presence of caustics and multiple images), so we can assume that in this case $n_r \sim1$.
Following the same procedure then used in Armstrong et al. we estimate 
$C_n^2 \sim 20$ m$^{-6.67}$ and $P_{\delta n_e} \sim 7 \times 10^{38}$ m$^{-3}$ at $l=6 \times 10^{10}$~m. 
If we apply the same calculation on PKS 1939-315 \citep{Bannister16}, we get
$C_n^2 \sim 2\times10^{-2}$ m$^{-6.67}$.
The $C_n^2$ values for both sources are higher (by a factor of $\sim10^4$ and $\sim10$, respectively) relative to Armstrong et al.'s $C_n^2=10^{-3}$ m$^{-6.67}$. However, 
these values are not hugely off when considering the dispersion of the Big Power Law relation.
While these are illustrative examples, they show that a turbulent origin of ESEs cannot be ruled out.

{\bf Theory.} While overpressure is not a problem for ``normal" pulsar scintillation, the evidence for structure on scales as small as 100 km provides a fascinating laboratory for magnetized
plasma turbulence from fluid to kinetic scales. While the solar wind is also such a laboratory, and one that can be probed {\textit{in situ}}, the wind is less collisional and known to be
dominated by supersonic flow. At the smallest scales at which structure is observed, the electrons and ions are largely decoupled and we are in the regime of electron fluid turbulence
\citep{Biskamp99,Cho2009}, which is known to develop intense filamentary structures.
Whether such turbulence has properties compatible with TSIS, and what power source is necessary to sustain it, have yet to be determined. For more
extreme structures, a theory for the formation of ``plasma lenses" is sorely needed, both to
complete the concept and to constrain the models (e.g. Gaussian vs power law slopes). The multiple sheets structures posited by
\cite{Romani87} should also be amenable to numerical simulation.













\section{Future advances, observational and theoretical}

The nature of turbulent dissipation processes, structures they produce and their local impact on dynamics and energy balance in the ISM remain at the frontier of astrophysics \citep{Hennebelle12}.
TSAS and TSIS probe spatial scales in the ballpark of what is currently expected for dissipation scales, therefore offering a unique peek at important and not-well understood physical processes with huge implications for many areas of astrophysics. 
Similarly, as we demonstrated in this review, the heating rates due to expansion of over-pressured TSAS and ESEs could be significant, meriting measurements of the filling factor of these structures across different interstellar environments.
In our view,  TSAS and TSIS are not only important for understanding turbulent dissipation, but  also 
offer a unique probe of the efficiency of stellar feedback as are likely produced by local turbulent enhancements in places like SNRs, stellar bubbles and winds.
In light of the importance of stellar feedback and turbulence for many areas of astrophysics, we think
that TSAS and TSIS deserve a major spotlight and attention. At the same time, demands for
increasing precision in pulsar timing for low frequency gravitational wave detection is spurring
efforts and building a new data base for modeling electron density fluctuation in the ISM.
Many upcoming advances in both observational facilities and theory will enable progress 
in the field.


Several upcoming radio facilities are promising a revolution in terms of angular resolution and sensitivity: the Australia Square Kilometre Array Pathfinder (ASKAP), MeerKAT, Next Generation VLA (ngVLA), the Square Kilometre Array (SKA). This will provide excellent opportunities for interferometric imaging of HI absorption, at high resolution, 
toward extended radio sources, e.g. SNRs. The goal is to measure the power spectrum of $\Delta \tau$ variations over a continuous range of scales from $\sim10$ pc all the way to $\sim10$ AU to test the turbulent origin of TSAS for a statistically significant  and diverse sample of interstellar environments. 
Such experiments can probe the turbulent spectrum, but also improve understanding of the cut-off scales imposed by thermal conduction, ambipolar diffusion and other processes.
As a complement, 
the Five-hundred-meter Aperture Spherical radio Telescope
(FAST) will be a wonderful telescope for long-term monitoring of TSAS discovered via pulsar observations simultaneously at HI and OH frequencies. Long-term monitoring will build variations of $\Delta \tau$ over spatial scales and reveal a roadmap of TSAS evolution with time.
The ongoing revolution in the integral field spectrograph (IFU)
astronomy will similarly provide optical absorption spectra simultaneously against hundreds of stars in clusters, enabling essentially 2D coverage and spatial power spectrum calculations 
using many species, which are currently not possible.

Another advantage some of these telescopes have is a 
large field of view (e.g. ASKAP, MeerKAT, SKA). 
Surveys under planning \citep{Dickey13,McClure-Griffiths15}
will provide close to a million Galactic HI absorption spectra (SKA). This will provide rich constraints on the range of spin temperatures found and their  spatial variations with interstellar environments, probing the dependence of heating and cooling, and thermal pressure, on the local environment. 
For understanding TSAS/ESE internal structure ALMA searches for CO, HCO$^+$, CH$^+$ and SiO will be essential to understand cooling and a possible shock-related origin of these structures. Similarly, shock tracers can probe the immediate surrounds of TSAS; if TSAS is expanding into the CNM strong shocks are expected.
Also, the James Webb Space Telescope (JWST) will be a wonderful instrument for measuring H$_2$ vibrational transitions in infrared, further constraining internal structure of both TSAS and ESEs.

Due to the ``searchlight'' nature of absorption profiles, exploring the temporal variability of absorption profiles will become abundantly more possible in the future. For example, ASKAP can stare at the Magellanic Clouds and in 50-200 hours of telescope time get an instantaneous  sample of hundreds of HI absorption lines. A repeated experiment over several time epochs will provide hundreds of $\Delta \tau$ probes. This is much cheaper in terms of telescope time than what is doable currently. Similarly, MeerKAT and SKA observations at different epochs will easily provide thousands of $\Delta \tau$ measurements.

As we highlighted earlier, pulsars have a unique ability to probe neutral, ionized and molecular media over exactly the same line of sight. But dedicated experiments are required. One potentially interesting way to do this is by adding spectral-line observations to pulsar timing measurements for the sample of sources observed by the North American Nanohertz Observatory for Gravitational Waves (NANOGrav) collaboration.
These pulsars are being monitored frequently to search for gravitational waves, however provide observations of DM over many time epochs that represent a goldmine for monitoring electron density structure in the ISM.

One of the key science goals of one of the several survey science projects accepted for ASKAP, VAST -- Variables and Slow Transients - is to determine the origin and nature
of structures responsible for ESEs.  ASKAP
will allow moderate sensitivity but wide field of view surveys (30 deg$^2$ in a single pointing)
enabling fast surveys and searches for transient sources, Murphy et al. 2013.
For example, Fiedler et al. (1994) searched about 600 sources in total for ESEs, VAST will be able to
cover close to 180,000 source-years. This will help to constrain the frequency and spatial distribution of refractive 
lenses, and possible associations with Galactic structures. In addition, real-time follow-up monitoring will
be provided of identified ESE sources -- this is crucial for probing the possible connection with
neutral structures, and to constrain magnetic field strength. 

There are frontiers on the theory side as well. Advances in understanding magnetized turbulence down to kinetic scales can and should be applied to the ISM. As challenging as this is on its own, accounting properly for factors such as  realistic heating and cooling processes, chemical reactions, suppression of thermal conduction and viscosity perpendicular to the ambient magnetic field, and the presence of cosmic rays are all essential for understanding the spectrum, degree of spatial intermittency, and
nature of dissipation. In order to connect with observations, and make predictions, it will be necessary to consider radiation transport through the medium as well.
Fortunately, with computational power increasing on rapidly, numerical simulations covering an
extended range of scales that reaches down to AU scales are becoming more feasible, and it is allowing  comprehensive modeling of bubble walls and supernova shells with unprecedented fidelity.



\section*{DISCLOSURE STATEMENT}
If the authors have noting to disclose, the following statement will be used: The authors are 
not aware of any affiliations, memberships, funding, or financial holdings that
might be perceived as affecting the objectivity of this review. 

\section*{ACKNOWLEDGMENTS}
We are grateful for many inspiring discussions and thought provoking
comments from the following colleagues, which have greatly improved the paper:
Haley Bignall, Avinash Deshpande, Miller Goss, Carl Heiles, David Kaplan, Joe Lazio, Chris McKee, Dave Meyer, Steve Spangler, Jacco van Loon, Joel Weisberg, and Mark Wolfire. 
We also appreciate and thank the following undergraduate students at UW for their help with the manuscript. Delano Yoder assisted in collecting information for Tables 1 and 2. Sam Szotkowski compared heating and cooling rates from several publications, and Eowyn Liu assisted in the reference compilation. We thank Roselyn Lowe-Webb at Annual Reviews for her patience with the completion of this manuscript, as well as help with the last-minute LaTeX questions.
EGZ acknowledges support from the University of Wisconsin-Madison and the hospitality of the University of Chicago, where part of this review was written.
SS acknowledges support from the National Science Foundation Early Career Award AST-1056780 and the 
 University of Wisconsin–Madison Office of the Vice Chancellor for Research and Graduate 
 Education Vilas Associate Award.

%
\section*{LITERATURE\ CITED}


\bibliographystyle{ar-style2}
\bibliography{myref_eowyn}

\end{document}